\documentclass[12pt,draftcls, peerreview, onecolumn]{IEEEtran/IEEEtran}
\usepackage{cite}
\usepackage{graphicx}
\usepackage{url}
\usepackage[centertags]{amsmath}
\usepackage{amsfonts}
\usepackage{amsthm}
\usepackage{multirow}
\newcommand{\beq}{\begin{equation}}
\newcommand{\eeq}{\end{equation}}
\newcommand{\beqn}{\begin{align}}
\newcommand{\eeqn}{\end{align}}

\begin{document}

\title{Secure Transmission in Multi-Cell Massive MIMO Systems}

\author{\IEEEauthorblockN{Jun~Zhu, Robert~Schober, and Vijay~K.~Bhargava}\\
\IEEEauthorblockA{The University of British Columbia\\
Email: \textit{\{zhujun,~rschober,~vijayb\}@ece.ubc.ca}}
\thanks{This work was presented in part at IEEE Globecom, Atlanta, December 2013.}}
\IEEEoverridecommandlockouts

\setcounter{page}{1}

\maketitle

\begin{abstract}
In this paper, we consider physical layer security provisioning in multi-cell massive multiple-input multiple-output (MIMO) systems. Specifically, we consider secure downlink transmission in a multi-cell massive MIMO system with
matched-filter precoding and artificial noise (AN) generation at the base station (BS) in the presence of a passive multi-antenna eavesdropper. We investigate the resulting achievable ergodic secrecy rate and the secrecy outage
probability for the cases of perfect training and pilot contamination. Thereby, we consider two different AN shaping matrices, namely, the conventional AN shaping matrix, where the AN is transmitted in the null space of the matrix
formed by all user channels, and a random AN shaping matrix, which avoids the complexity associated with finding the null space of a large matrix.
Our analytical and numerical results reveal that in multi-cell massive MIMO systems employing matched-filter precoding (1) AN generation is required to achieve a positive ergodic secrecy rate if the user and the eavesdropper
experience the same path-loss, (2) even with AN generation secure transmission may not be possible if the number of eavesdropper antennas is too large and not enough power is allocated to channel estimation, (3) for a given
fraction of power allocated to AN and a given number of users, in case of pilot contamination, the ergodic secrecy rate is not a monotonically increasing function of the number of BS antennas, and (4) random AN shaping matrices
provide a favourable performance/complexity tradeoff and are an attractive alternative to conventional AN shaping matrices.
\end{abstract}


\section{Introduction}
Security is a vital issue in wireless networks due to the broadcast nature of the medium. Traditionally, security has been achieved through cryptographic encryption implemented at the application layer. This approach is based on certain
assumptions regarding computational complexity, and is thereby potentially vulnerable \cite{sec_survey}. As a complement to cryptographic methods, physical layer security has drawn significant research and industrial interest recently.
The pioneering work on physical layer security in \cite{wiretap} considered the classical three-terminal network consisting of a transmitter (Alice), an intended receiver (Bob), and an eavesdropper (Eve). It was shown in \cite{wiretap} that
a source-destination pair can exchange perfectly secure messages with a positive rate as long as the desired receiver enjoys better channel conditions than the eavesdropper(s). More recent studies have considered physical layer security
provisioning in multi-antenna multiuser networks \cite{khisti}-\cite{GEYRC12}. Although the secrecy capacity region for multiuser networks remains an
open problem, it is interesting to investigate the achievable secrecy rates of such networks for certain practical transmission strategies.
Eavesdroppers are typically passive so as to hide their existence, and thus their channel state information (CSI) cannot be obtained by Alice \cite{masksec}. In this case, multiple transmit antennas can be exploited to enhance secrecy by
simultaneously transmitting both the information-bearing signal and artificial noise (AN) \cite{negi}. Specifically, precoding is used to make the AN invisible to Bob while degrading the decoding performance of possibly present Eves
\cite{masksec}-\cite{robsec}. For the case of imperfect channel estimation, robust beamforming designs were reported in \cite{masksec,robsec}.

Recently, a new promising design approach for cellular networks, known as massive or large-scale multiple-input multiple-output (MIMO), has been proposed \cite{survey, noncooperative}, where base station (BS) antenna
arrays are equipped with an order of magnitude more elements than what is used in current systems, i.e., a hundred antennas or more. Massive MIMO enjoys all the benefits of conventional multiuser MIMO,
such as improved data rate, reliability and reduced interference, but at a much larger scale and with simple linear precoding/detection schemes. In fact, massive MIMO employing simple matched-filter precoding/combining
enables large gains in bandwidth and/or power efficiency compared to conventional MIMO systems \cite{noncooperative,energyspectraleff} as the effects of noise and interference vanish completely in the limit of an infinite number
of antennas. Furthermore, in time-division duplex (TDD) systems, channel reciprocity can be exploited to estimate the downlink channels via uplink training such that the resulting overhead scales linearly with the number of users but
is independent of the number of BS antennas \cite{training}. However, if the pilot sequences employed in different cells are not orthogonal, so-called pilot contamination impairs the channel estimates and limits the achievable
information rates in massive MIMO systems \cite{noncooperative,pilotcontam}.

Massive MIMO systems offer an abundance of BS antennas, while multiple transmit antennas can be exploited for secrecy enhancement. Therefore, the combination of both concepts seems natural and
promising, which is the main motivation for the present work. However, several new issues arise for physical layer security provisioning in multi-cell massive MIMO systems that are not present for conventional MIMO systems
\cite{sec_survey}-\cite{robsec}. For example, pilot contamination is unique to massive MIMO systems and we study its effect on the ergodic secrecy rate and the secrecy outage probability. Furthermore, for the user data,
matched-filter precoding is usually adopted in massive MIMO systems \cite{survey, noncooperative}, since the matrix inversion needed for the schemes used in conventional MIMO, such as regularized zero-forcing (ZF) and
minimum mean squared error (MMSE) precoding, is considered to be computationally too expensive for the large matrices typical for massive MIMO. Similarly, whereas in conventional MIMO systems the AN is transmitted in the
null space of the channel matrix \cite{negi}, the complexity associated with computing the null space may not be affordable in case of massive MIMO and simpler AN shaping methods may be needed. Finally, unlike most of the
related work \cite{sec_survey}-\cite{robsec}, we consider a multi-cell setting where not only the data signals cause inter-cell interference but also the AN, which has to be carefully taken into account for system design.

In this paper, we study  secure downlink transmission in multi-cell massive MIMO systems in the presence of a multi-antenna eavesdropper, which attempts to intercept the signal intended for one of the users. To arrive at
an achievable secrecy rate for this user, we assume that the eavesdropper can acquire perfect knowledge of the channel state information (CSI) of all user data channels and is able to cancel all interfering user signals. Under this assumption,
we derive tight lower bounds for the ergodic secrecy rate and tight upper bounds for the secrecy outage probability for the cases of perfect training and pilot contamination. The derived bounds are in closed form and provide significant
insight for system design. In particular, the obtained results allow us to predict under what conditions (i.e., for what number of BS antennas, eavesdropper antennas, users, path-loss, number of cells, and pilot powers) a positive secrecy rate is possible.
Furthermore, we show that employing random AN shaping matrices is an attractive low-complexity option for massive MIMO systems. We also derive a closed-form expression for the fraction
of transmit power that should be optimally allocated to AN and show that, for a given number of BS antennas, this fraction increases with the number of eavesdropper antennas and decreases with the number of users in the system.

\textit{Notations:} Subscripts $T$ and $H$ stand for the transpose and the conjugate transpose, respectively. ${\bf I}_N$ and ${\bf 0}_N$ denote the $N$-dimensional identity matrix and the all-zero column vector of length $N$, respectively.
The expectation operation, variance operation, and Euclidean norm are denoted by $\mathbb{E}[\cdot]$, ${\rm var}[\cdot]$, and $\|\cdot\|^2$, respectively. $\mathbb{C}^{m\times n}$ represents the space of all $m \times n$ matrices with
complex-valued elements. Furthermore, $\overline{X}$ and $\underline{X}$ denote an upper bound and a lower bound for $X$, respectively, i.e., $\underline{X}\le X\le\overline{X}$. Finally, we use ${\bf x} \sim \mathbb{CN}({\bf 0}_N,
\boldsymbol{\Sigma})$ to denote a circularly symmetric complex Gaussian vector ${\bf x}\in \mathbb{C}^{N\times 1}$ with zero mean and covariance matrix $\boldsymbol{\Sigma}$, and $x\sim \chi^2_{n}$ means that $\sqrt{2}x$ is a
chi-square random variable with $n$ degrees of freedom.

\section{System Model \label{s2}}
In this section, we introduce the channel model, the channel estimation scheme, the transmission format, and two AN shaping matrix designs
for the considered secure multi-cell massive MIMO system. For convenience, the most important variables used in this paper are defined in Table \ref{table1}.
\subsection{System and Channel Models}
  \begin{figure}
  \centering
    \includegraphics[width=4in]{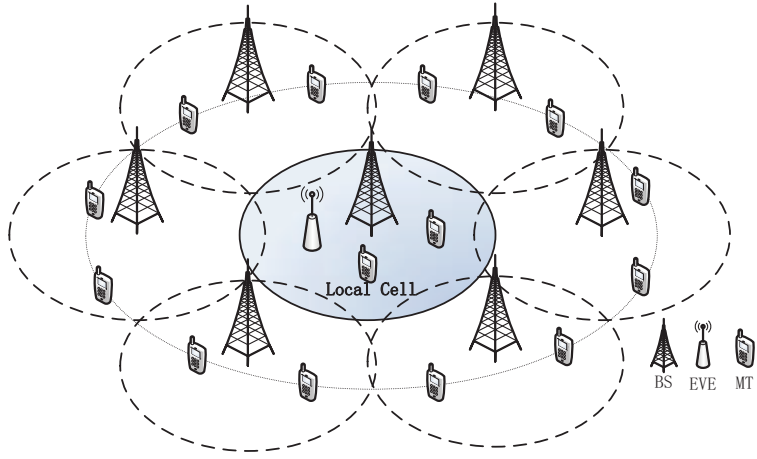}\\
    \caption{Multi-cell massive MIMO system in the presence of a multi-antenna eavesdropper. The shaded cell is the local cell. The MTs in the local cell suffer from the inter-cell interference caused by data and AN transmission in the
    six adjacent cells.}\label{Figsys}
  \end{figure}
In this paper, we consider a flat-fading multi-cell system consisting of $M$ cells, as depicted in Fig.~\ref{Figsys}. Each cell comprises an $N_t$-antenna BS and $K$ single-antenna mobile terminals (MTs)\footnote{We note that
the results derived in this paper can be easily extended to multi-antenna MTs if the BS transmits one independent data stream per MT receive antenna and receive combining is not performed at the MTs. In this case, each MT
receive antenna can be treated as one (virtual) MT and the results derived in this paper are applicable. For example, the secrecy rate of a multi-antenna MT can be obtained by summing up the secrecy rates of its
receive antennas.}. The $n^{\rm th}$ cell, $n \in \{1,\ldots,M\}$, is the local cell (the shaded area in Fig. \ref{Figsys}). An eavesdropper equipped with $N_e$ antennas (equivalent to $N_e$ cooperative single-antenna
eavesdroppers) is located in the local cell of the considered multi-cell region. The eavesdropper is passive and seeks to recover the information transmitted to the $k^{\rm th}$ MT in the local cell. Let ${\bf h}_{mk} \in \mathbb{C}^{1 \times N_t}$ and
${\bf H}^{\rm eve}_{m} \in \mathbb{C}^{N_e \times N_t}$ denote the channel between the $m^{\rm th}$ BS, $m=1,\ldots, M$, and the $k^{\rm th}$ MT in the local cell and the channel between the $m^{\rm th}$ BS and the eavesdropper, respectively.
${\bf h}_{mk}=\sqrt{l_{mk}} \tilde{\bf h}_{mk}$ comprises the path-loss, $l_{mk}$, and the small-scale fading vector, $\tilde{\bf h}_{mk}\sim  \mathbb{CN}({\bf 0}_{N_t}^T,{\bf I}_{N_t})$. Similarly, we model the eavesdropper channel as
${\bf H}^{\rm eve}_{m} =\sqrt{l_m^{\rm eve}}  \tilde{\bf H}^{\rm eve}_{m}$, where $l_m^{\rm eve}$ and $\tilde{\bf H}^{\rm eve}_{m}$ denote the path-loss and small-scale fading components, respectively. The elements of $\tilde{\bf H}^{\rm eve}_{m}$
are modeled as independent and identically distributed (i.i.d.) Gaussian random variables (r.v.s) with zero mean and unit variance.

\begin{table}
\caption{Summary of most important variables used in this paper.}
 \label{table1}
   \centering
   \begin{tabular}{|l|l|}
         \hline
         \multicolumn{1}{|l|}{\bf{Symbols}} & \bf{Description}\\
         \hline
         $M$ & Number of cells\\
         $N_t$ & Number of BS antennas\\
         $N_e$ & Number of eavesdropper antennas\\
         $\alpha$ & Ratio of the number of eavesdropper antennas and the number of BS antennas\\
         $K$ & Number of MTs in each cell\\
         $\beta$ & Ratio of the number of MTs in each cell and the number of BS antennas\\
         $l_{mk}$ & Path-loss between the $m^{\rm th}$ BS and the $k^{\rm th}$ MT in the local cell\\
         $l^{\rm eve}_m$ & Path-loss between the $m^{\rm th}$ BS and the eavesdropper\\
         $\rho$ & Inter-cell interference factor\\
          $P$ & Total transmit power\\
         $\phi$ & Power allocation factor between signal transmission and AN generation\\
         $p_\tau$ & Power of pilot symbol\\
         $\tau$ & Pilot sequence length\\
         $a$ &  Parameter ($a=(M-1)\rho+1$)\\
         $b$  & Parameter ($b=(M-1)\rho+1/P$)\\
         $c$ & Parameter ($c=(M-1)\rho^2+1$)\\
         $\zeta$ & Parameter ($\zeta= a \beta/\alpha-\beta c/[a(1-\beta)]$)\\
         $\lambda$ & Measure for quality of channel estimates ($\lambda=p_\tau\tau/(1+p_\tau\tau a)$)\\
          $\alpha_{\rm sec}$ & A positive secrecy rate is possible only if $\alpha< \alpha_{\rm sec}$\\
         $\tilde{\bf h}_{mk}$ & Small-scale fading vector between the $m^{\rm th}$ BS and the $k^{\rm th}$ MT in the local cell\\
         ${\bf h}_{mk}$ & Channel vector between the $m^{\rm th}$ BS and the $k^{\rm th}$ MT in the local cell\\
         $\hat{\bf h}_{mk}$ & Estimate of the channel vector between the $m^{\rm th}$ BS and the $k^{\rm th}$ MT in the local cell\\
         $\tilde{\bf H}^{\rm eve}_m$ & Small-scale fading matrix between the $m^{\rm th}$ BS and the eavesdropper\\
                 ${\bf H}^{\rm eve}_m$ & Channel matrix between the $m^{\rm th}$ BS and the eavesdropper\\
         $\gamma_{nk}$ & Received SINR at the $k^{\rm th}$ MT in the local cell\\
         $\gamma_{\rm eve}$ & Received SINR at the eavesdropper\\
         $R_{nk}$ & Achievable ergodic rate of the $k^{\rm th}$ MT in the local cell\\
         $C^{\rm eve}_{nk}$ & Ergodic capacity of the eavesdropper seeking to decode the signal of the $k^{\rm th}$ MT in the local cell\\\
         $R^{\rm sec}_{nk}$ & Ergodic secrecy rate of the $k^{\rm th}$ MT in the local cell\\
         $\varepsilon_{\rm out}$ & Secrecy outage probability of the $k^{\rm th}$ MT in the local cell\\
          \hline
     \end{tabular}
 \end{table}

\subsection{Uplink Training and Channel Estimation\label{s2a}}
In this paper, we assume that the BSs are perfectly synchronized and operate in the TDD mode with universal frequency reuse. Furthermore, we assume that the path-losses between all users in the system and the local BS,
$l_{mk}$, $m=1,\ldots, M$, $k=1,\ldots K$, are known at the local BS, whereas the small-scale fading vectors $\tilde{\bf h}_{mk}$, $m=1,\ldots, M$, $k=1,\ldots K$, are not known and the local BS estimates only the
small-scale fading vectors of the MTs within the local cell. These assumptions are motivated by the fact that the path-losses change on a much slower time scale than the small-scale fading vectors, and thus, their estimation creates
a comparatively low overhead.

The local BS estimates the downlink CSI of all MTs, $\tilde{\bf h}_{nk}$, $k=1,\ldots, K$, by exploiting reverse training and channel reciprocity \cite{survey}-\cite{pilotcontam}. We consider two scenarios: Perfect training and imperfect training
which results in pilot contamination \cite{pilotcontam}. In the former case, all $MK$ MTs in the system emit orthogonal pilot sequences in the training phase having a sufficiently large pilot power $p_\tau$ such that $\hat{\bf h}_{nk}=\tilde{\bf h}_{nk}$,
$k=1,\ldots, N_t$, can be assumed, where $\hat{\bf h}_{nk}$ denotes the estimated channel in the local cell. In the latter case, the $K$ pilot sequences used in a cell are still orthogonal but all cells use the same pilot sequences. Let $\sqrt{\tau} \boldsymbol{\omega}_{k} \in \mathbb{C}^{\tau \times 1}$ denote the pilot sequence of length $\tau$ transmitted by the $k^{\rm th}$ MT in each cell in the training phase, where $\boldsymbol{\omega}^H_{k}\boldsymbol{\omega}_{k}=1$ and $\boldsymbol{\omega}^H_{k}\boldsymbol{\omega}_{j}=0$, $\forall , j, k=1,\ldots,K$, $k \neq j$. Assuming perfect synchronization, the training signal received at the local BS, ${\bf Y}_{n}^{\text{pilot}} \in \mathbb{C}^{\tau \times N_t}$, can be expressed as
\begin{equation}
{\bf Y}_{n}^{\text{pilot}}=\sum_{m=1}^M \sum_{k=1}^K \sqrt{p_{\tau} \tau l_{mk}} \boldsymbol {\omega}_{k} \tilde{\bf h}_{mk}+{\bf N}_{n},
\end{equation}
where ${\bf N}_n \in \mathbb{C}^{\tau \times N_t}$ is a Gaussian noise matrix having zero mean, unit variance elements. Assuming MMSE channel estimation \cite{pilotcontam}, \cite{uldl}, the estimate
of $\tilde{\bf h}_{nk}$ given ${\bf Y}_{n}^{\text{pilot}}$ is obtained as
\begin{eqnarray}
\label{est}
 \hspace{0 cm} \nonumber \hat{{\bf h}}_{nk}&=&\sqrt{p_{\tau} \tau l_{nk}} \boldsymbol{\omega}^H_{k}\left({\bf I}_{\tau}+\boldsymbol{\omega}_{k}\left(p_{\tau} \tau \sum_{m=1}^M l_{mk}\right)\boldsymbol{\omega}^H_{k}\right)^{-1} {\bf Y}_{n}^{\text{pilot}}
 =\frac{\sqrt{p_{\tau} \tau l_{nk}}}{1+p_{\tau} \tau \sum_{m=1}^M l_{mk}}\boldsymbol{\omega}^H_{k}{\bf Y}_{n}^{\text{pilot}}\\
\hspace{0 cm}& =&\frac{\sqrt{p_{\tau} \tau l_{nk}}}{1+p_{\tau} \tau \sum_{m=1}^M l_{mk}}\sum_{m=1}^M \sqrt{p_{\tau} \tau} \tilde{\bf h}_{mk}+\frac{\sqrt{p_{\tau} \tau l_{nk}}}{1+p_{\tau} \tau \sum_{m=1}^M l_{mk}} \boldsymbol{\omega}^H_{k} {\bf N}_n.
\end{eqnarray}
For MMSE estimation, we can express the channel as $\tilde{\bf h}_{nk}=\hat{{\bf h}}_{nk}+{\bf e}_{nk}$, where the estimate $\hat{{\bf h}}_{nk}$ and the estimation error ${\bf e}_{nk}\in \mathbb{C}^{1 \times N_t}$ are mutually independent.
Hence, considering (\ref{est}) we can statistically characterize $\hat{{\bf h}}_{nk}$ and ${\bf e}_{nk}$ as $\hat{{\bf h}}_{nk} \sim \mathbb{CN}\left({\bf 0}_{N_t}^T, \frac{p_\tau \tau l_{nk}}{1+p_\tau \tau \sum_{m=1}^M l_{mk}}{\bf I}_{N_t}\right)$ and
${\bf e}_{nk} \sim \mathbb{CN}\left({\bf 0}_{N_t}^T, \frac{1+p_\tau \tau \sum_{m \neq n}l_{mk}}{1+p_\tau \tau \sum_{m=1}^M l_{mk}}{\bf I}_{N_t}\right)$, respectively.

We note that in order to be able to find the required numbers of orthogonal pilot sequences, pilot sequence lengths of $\tau\ge MK$ and $\tau\ge K$ are required for the cases of perfect training and pilot contamination, respectively.
Furthermore, we note that the eavesdropper could emit his own pilot symbols to impair the channel estimates obtained at the BS to improve his ability to decode the MTs' signals during downlink transmission \cite{ZMH12}. However, this would also increase the
chance that the presence of the eavesdropper is detected by the BS \cite{KZWO13}. Therefore, in this paper, we assume the eavesdropper is purely passive and leave the study of active eavesdroppers in massive MIMO systems for future work.

\subsection{Downlink Data Transmission\label{s2b}}
In the local cell, the BS intends to transmit a confidential signal $s_{nk}$ to the $k^{\rm th}$ MT. The signal vector for the $K$ MTs is denoted by ${\bf s}_n=\big[s_{n1},\ldots,s_{nK}\big]^T \in \mathbb{C}^{K \times 1}$ with
$\mathbb{E}[{\bf s}_n {\bf s}_n^H]={\bf I}_K$. Each signal vector ${\bf s}_{n}$ is multiplied by a transmit beamforming matrix, ${\bf W}_n=[{\bf w}_{n1},\ldots,{\bf w}_{nk},\ldots,{\bf w}_{nK}]\in \mathbb{C}^{N_t \times K}$, before
transmission. As typical for massive MIMO systems, we adopt simple matched-filter precoding, i.e., ${\bf w}_{nk}={\hat{\bf h}}_{nk}^H/\|{\hat{\bf h}}_{nk}\|$ \cite{noncooperative},\cite{pilotcontam}, since the matrix inversion
required for ZF and MMSE precoding is computationally too expensive for the large number of users and antenna elements that are typical for massive MIMO systems. Furthermore, we assume that the eavesdropper's
CSI is not available at the local BS. Hence, assuming that there are $K<N_t$ MTs, the BS may use the remaining $N_t-K$ degrees of freedom offered by the $N_t$ transmit antennas for
emission of AN to degrade the eavesdropper's ability to decode the data intended for the MTs \cite{masksec,negi,robsec}. The AN vector, ${\bf z}_n =[z_{n1},\ldots,z_{n(N_t-K)}]^T\sim \mathbb{CN}({\bf 0}_{N_t-K},{\bf I}_{N_t-K})$, is multiplied by an AN
shaping matrix ${\bf V}_n=[{\bf v}_{n1},\ldots,{\bf v}_{ni},\ldots,{\bf v}_{n (N_t-K)}]\in \mathbb{C}^{N_t \times (N_t-K)}$ with $\|{\bf v}_{ni}\|=1$, $i=1,\ldots,N_t-K$. The considered choices for the AN shaping matrix will be discussed in
the next subsection. The signal vector transmitted by the local BS is given by
\begin{equation}
{\bf x}_n=\sqrt{p} {\bf W}_n {\bf s}_n+\sqrt{q} {\bf V}_n{\bf z}_n=\sum_{k=1}^K \sqrt{p}{\bf w}_{nk} s_{nk}+\sum_{i=1}^{N_t-K} \hspace*{-2mm}\sqrt{q}{\bf v}_{ni} z_{ni},
\end{equation}
where $p$ and $q$ denote the transmit power allocated to each MT and each AN signal, respectively, i.e., for simplicity, we assume uniform power allocation across users and AN signals, respectively. Let the total transmit power be denoted
by $P$. Then, $p$ and $q$ can be represented as $p=\frac{\phi P}{K}$ and $q=\frac{(1-\phi)P}{N_t-K}$, respectively, where the power allocation factor $\phi$, $0<\phi \le 1$, strikes a power balance between the
information-bearing signal and the AN.

The $M-1$ cells adjacent to the local cell transmit their own signals and AN. In this work, in order to be able to gain some fundamental insights, we assume that all cells employ identical values for $p$ and $q$ as well as  $\phi$.
Accordingly, the received signals at the $k^{\rm th}$ MT in the local cell, $y_{nk}$, and at the eavesdropper, ${\bf y}_{\rm eve}$, are given by
\begin{eqnarray}
y_{nk}&=&\sqrt{p}{\bf h}_{nk}{\bf w}_{nk}  s_{nk}+\sum_{\{m,l\} \neq \{n,k\}}  \sqrt{p}{\bf h}_{mk}{\bf w}_{ml} s_{ml} +\sum_{m=1}^M \sqrt{q} {\bf h}_{mk} {\bf V}_m{\bf z}_m+n_{nk}
\label{x1}
\\
{\bf y}_{\rm eve}&=& \sqrt{p} \sum_{m=1}^M  {\bf H}^{\rm eve}_{m} {\bf W}_m {\bf s}_m+\sqrt{q}\sum_{m=1}^M  {\bf H}^{\rm eve}_{m} {\bf V}_m{\bf z}_m+{\bf n}_{\rm eve},
\label{x2}
\end{eqnarray}
where $n_{nk}\sim  \mathbb{CN}(0,\sigma_{nk}^2)$ and ${\bf n}_{\rm eve}\sim  \mathbb{CN}({\bf 0}_{N_e}, \sigma_{\rm eve}^2{\bf I}_{N_e})$ are the Gaussian noises at the $k^{\rm th}$ MT and at the eavesdropper, respectively.
The first term on the right hand side of (\ref{x1}) is the signal intended for the $k^{\rm th}$ MT in the local cell with effective channel gain $\sqrt{p}{\bf h}_{nk}{\bf w}_{nk}$, which is assumed to be perfectly known at the $k^{\rm th}$ MT
in the local cell. The second and the third terms on the right hand side of (\ref{x1}) represent intra-cell/inter-cell interference and AN leakage, respectively. On the other hand, the eavesdropper observes
an $MN_t\times N_e$ MIMO channel comprising $K$ local user signals, $(M-1)K$ out-of-cell user signals, $N_t-K$ local cell AN signals, and $(N_t-K)(M-1)$ out-of-cell AN signals. In order to obtain a lower bound on the achievable secrecy
rate, we assume that the eavesdropper can acquire perfect knowledge of the effective channels of all MTs, i.e., ${\bf H}^{\rm eve}_{m} {\bf w}_{mk},\forall m,k$. We note however that this is a quite pessimistic assumption because the
uplink training performed in massive MIMO \cite{pilotcontam} makes it difficult for the eavesdropper to perform accurate channel estimation.
\subsection{Design of AN Shaping Matrix ${\bf V}_n$\label{s2c}}
In this paper, we consider two different designs for the AN shaping matrix ${\bf V}_n$.

\textbf{Null-space method:} For conventional (non-massive) MIMO, ${\bf V}_n$ is usually chosen to lie in the null space of the estimated channel, $\hat{\bf h}_{nk}$, i.e., $\hat{\bf h}_{nk} {\bf V}_n={\bf 0}_{N_t-K}^T$, $k=1,\ldots,K$, which is
possible as long as $N_t > K$ holds \cite{negi}. We refer to this method as $\mathcal{N}$ in the following. If perfect CSI is available, i.e., $\hat{\bf h}_{nk}= \tilde{\bf h}_{nk} ,$ the $\mathcal{N}$-method prevents impairment of the users in
the local cell by AN generated by the local BS. However, in case of pilot contamination, AN leakage to the users in the local cell is unavoidable. More importantly, for the large values of $N_t$ and $K$ typical for massive MIMO systems,
computation of the null space of $\hat{\bf h}_{nk}$, $k=1,\ldots,K$, is computationally expensive. This motivates the introduction of a simpler method for generation of the AN shaping matrix.

\textbf{Random method:} In this case, the columns of ${\bf V}_n$ are mutually independent random vectors. We refer to this method as $\mathcal{R}$ in the following. Here, we construct the columns of ${\bf V}_n$ as ${\bf v}_{ni}=\tilde{\bf v}_{ni}/
\|\tilde{\bf v}_{ni}\|$, where the $\tilde{\bf v}_{ni}$, $i=1,\ldots,N_t-K$, are mutually independent Gaussian random vectors. Note that the $\mathcal{R}$-method does not even attempt to avoid AN leakage to the users in the local cell.
However, it may still improve the ergodic secrecy rate as the precoding vector for the desired user signal, ${\bf w}_{nk}$, is correlated with the user channel, $\tilde{\bf h}_{nk}$, whereas the columns of the AN shaping matrix are not
correlated with the user channel.

Our results in Sections IV-VI reveal that although the $\mathcal{N}$-method always achieves a better performance than the $\mathcal{R}$-method, if pilot contamination and inter-cell interference are significant, the
performance differences between both schemes are small. This makes the $\mathcal{R}$-method an attractive alternative for massive MIMO systems due to its simplicity.

\section{Achievable Ergodic Secrecy Rate Analysis\label{sx}}
In this section, we first show that the achievable ergodic secrecy rate of the $k^{\rm th}$ MT in the local cell can be expressed as the difference between the achievable ergodic rate of the MT and the ergodic capacity of the eavesdropper.
Subsequently, we provide a simple lower bound on the achievable ergodic rate of the MT, a closed-form expression for the ergodic capacity of the eavesdropper, and a simple and tight upper bound for the ergodic capacity
of the eavesdropper. The results derived in this section are valid for both perfect training and pilot contamination as well as for both AN shaping matrix designs.
For convenience, we define the ratio of the number of eavesdropper antennas and the number of BS antennas as $\alpha=N_e/N_t$, and the ratio of the number of users and the number of BS antennas as $\beta= K/N_t $. In the
following, we are interested in the asymptotic regime where $N_t\to\infty$ but $\alpha$ and $\beta$ are constant.
\subsection{Achievable Ergodic Secrecy Rate\label{s2d0}}
The ergodic secrecy rate is an appropriate performance measure if delays can be afforded and coding over many independent channel realizations (i.e., over many coherence intervals) is possible \cite{zhou}. Considering the
$k^{\rm th}$ MT in the local cell, the considered channel is an instance of a multiple-input, single-output, multiple eavesdropper (MISOME) wiretap channel \cite{khisti}. In the following lemma, we provide an expression for an achievable
ergodic secrecy rate of the $k^{\rm th}$ MT in the local cell.

\textit{Lemma 1:} An achievable ergodic secrecy rate of the $k^{\rm th}$ MT in the local cell is given by
\begin{equation}
\label{eq8}
R_{nk}^{\rm sec}=[R_{nk}-C_{nk}^{\rm eve}]^+,
\end{equation}
where $[x]^+=\max\{0, x\}$, $R_{nk}$ is an achievable ergodic rate of the $k^{\rm th}$ MT in the local cell, and $C_{nk}^{\rm eve}$ is the ergodic capacity between the local BS and the eavesdropper seeking to decode the
information of the $k^{\rm th}$ MT in the local cell. Thereby, it is assumed that the eavesdropper is able to cancel the received signals of all in-cell and out-of-cell MTs except the signal intended for the MT of interest, i.e.,
\begin{equation}
\label{Ceve}
C_{nk}^{\rm eve} = \mathbb{E}\bigg[\log_2 \left(1+p {\bf w}^H_{nk} {\bf H}^{{\rm eve}H}_{n} {\bf X}^{-1}{\bf H}^{\rm eve}_{n} {\bf w}_{nk}\right)\bigg],
\end{equation}
where ${\bf X}=q \sum_{m=1}^M{\bf V}^H_{m} {\bf H}^{{\rm eve}H}_{m} {\bf H}^{\rm eve}_{m} {\bf V}_{m}$ denotes the noise correlation matrix at the eavesdropper under the worst-case assumption that the
receiver noise is negligible, i.e., $\sigma_{\rm eve}^2\to 0$.
\begin{proof}
Please refer to Appendix A.
\end{proof}
Eq.~(\ref{eq8}) reveals that the achievable ergodic secrecy rate of the $k^{\rm th}$ MT in the local cell has the subtractive form typical for many wiretap channels \cite{sec_survey}-\cite{robsec}, i.e., it is the difference of an
achievable ergodic rate of the user of interest and the capacity of the eavesdropper. Before we analyze (\ref{eq8}) for perfect training and pilot contamination in Sections IV and V, respectively, we derive general expressions for
$R_{nk}$ and $C_{nk}^{\rm eve}$, which apply to both cases.
\subsection{Lower Bound on the Achievable User Rate\label{s2d}}
Based on (\ref{x1}) an achievable ergodic rate of the $k^{\rm th}$ MT in the local cell is given by
\begin{equation}
\label{rate}
R_{nk}=\mathbb{E}\left[\log_2 \left(1+\frac{|\sqrt{p} {\bf h}_{nk} {\bf w}_{nk}|^2}{\sum_{m=1}^M \sum_{i=1}^{N_t-K} |\sqrt{q}{\bf h}_{mk} {\bf v}_{mi}|^2+\sum_{\{m,l\} \neq \{n,k\}} |\sqrt{p} {\bf h}_{mk} {\bf w}_{ml}|^2
+\sigma_{nk}^2}\right)\right].
\end{equation}
Unfortunately, evaluating the expected value in (\ref{rate}) analytically is cumbersome. Therefore, we derive a lower bound on the achievable ergodic rate of the $k^{\rm th}$ MT in the local cell by following the same approach as in
\cite{pilotcontam}. In particular, we rewrite the received signal at the $k^{\rm th}$ MT in the local cell as
\begin{equation}
\label{eq7}
y_{nk}=\mathbb{E}[\sqrt{p}{\bf h}_{nk}{\bf w}_{nk}]s_{nk}+n'_{nk},
\end{equation}
where $n'_{nk}$ represents an effective noise, which is given by
\begin{equation}
n'_{nk}=\left(\sqrt{p}{\bf h}_{nk}{\bf w}_{nk}-\mathbb{E}[\sqrt{p}{\bf h}_{nk}{\bf w}_{nk}]\right)s_{nk}+\sum_{m=1}^M {\bf h}_{mk} \sqrt{q} {\bf V}_m{\bf z}_m+\sum_{\{m,l\} \neq \{n,k\}}  \sqrt{p}{\bf h}_{mk}{\bf w}_{ml} s_{ml}+n_{nk}.
\end{equation}
Eq.~(\ref{eq7}) can be interpreted as an equivalent single-input single-output channel with constant gain $\mathbb{E}[\sqrt{p}{\bf h}_{nk}{\bf w}_{nk}]$ and AWGN $n'_{nk}$. Hence, we can apply \textit{Theorem 1} in \cite{pilotcontam}
to obtain a computable lower bound for the achievable rate of the $k^{\rm th}$ MT in the local cell as $\underline{R}_{nk}=\log_2 (1+\gamma_{nk})\le R_{nk}$, where $\gamma_{nk}$ denotes the received signal-to-interference-plus-noise ratio (SINR)
\begin{equation}
\label{gammadl}
\gamma_{nk}=\frac{\overset{\mbox{desired signal}}{\overbrace{|\mathbb{E}[\sqrt{p} {\bf h}_{nk} {\bf w}_{nk}]|^2}}}{\underset{\mbox{signal leakage}}{\underbrace{{\rm var}[\sqrt{p} {\bf h}_{nk} {\bf w}_{nk}]}}+\underset{\mbox{AN leakage}}{\underbrace{\sum_{m=1}^M \sum_{i=1}^{N_t-K} \mathbb{E}[|\sqrt{q}{\bf h}_{mk} {\bf v}_{mi}|^2]}}+\underset{\mbox{intra- and inter-cell interference}}{\underbrace{\sum_{\{m,l\} \neq \{n,k\}} \mathbb{E}[|\sqrt{p} {\bf h}_{mk} {\bf w}_{ml}|^2]}}
+\sigma_{nk}^2}
\end{equation}
with ${\rm var}[\sqrt{p} {\bf h}_{nk} {\bf w}_{nk}]=\mathbb{E}[|\sqrt{p} {\bf h}_{nk} {\bf w}_{nk}-\mathbb{E}[\sqrt{p} {\bf h}_{nk} {\bf w}_{nk}]|^2]$. We note that the derived lower bound on the achievable rate is applicable to both AN shaping matrix designs
and the cases of perfect training and pilot contamination, respectively, cf.~Sections IV and V. The tightness of the lower bound will be confirmed by our results in Section VI.
\subsection{Ergodic Capacity of the Eavesdropper\label{s2e}}
In this section, we provide a closed-form expression for the ergodic capacity of the eavesdropper valid for both perfect training and pilot contamination. To gain more insight, we adopt a simplified path-loss model for the eavesdropper, i.e.,
the path-losses between the BSs and the eavesdropper are given by $l_{m}^{\rm eve}=1$ if $n=m$ and $l_{m}^{\rm eve}=\rho$ if $n\ne m$, i.e., the path-loss between the local BS and the eavesdropper is $1$ and the path-loss between the BSs of the
other cells and the eavesdropper is  $\rho \in [0,1]$.\footnote{We note that the simplified path-loss model is only adopted to reduce the number of parameters. The ergodic capacity and the ergodic secrecy rate can also be derived for the original
path-loss model in closed form. However, the resulting equations are more cumbersome and less insightful compared to those for the simplified model.} A similar simplified path-loss model was used in \cite{uldl} for the user channels. The
resulting ergodic secrecy capacity is summarized in the following theorem.

\textit{Theorem 1:} For $N_t\to\infty$ and both the $\mathcal{N}$ and the $\mathcal{R}$ AN shaping matrix designs, the ergodic capacity of the eavesdropper in (\ref{Ceve}) can be written as
\begin{equation}
\label{Cint}
C^{\rm eve}_{nk} = \frac{1}{\ln 2} \sum_{i=0}^{N_e-1} \lambda_i \times \frac{1}{\mu_0} \sum_{j=1}^2 \sum_{l=2}^{b_j} \omega_{jl}I(1/\mu_j,l),
\end{equation}
where $\lambda_i=\binom{M(N_t-K)}{i}$,  $\mu_0=\prod_{j=1}^2 \mu_j^{b_j}$,
\begin{equation}
\label{Fyy}
(\mu_j,b_j)=\begin{cases} (\eta,N_t-K), & j=1\\
(\rho \eta,(M-1)(N_t-K)), & j=2,
\end{cases}
\end{equation}
$\eta=q/p$,
\begin{equation}
\omega_{jl}=\frac{1}{(b_j-l)!} \frac{d^{b_j-l}}{dx^{b_j-l}} \left(\frac{x^i}{\prod_{s \neq j} (x+\frac{1}{\mu_s})^{b_s}}\right)\bigg|_{x=-\frac{1}{\mu_j}},
\end{equation}
and $I(a,n)=\int_0^{\infty}\frac{1}{(x+1)(x+a)^n}dx,~a,n>0$. A closed-form expression for $I(\cdot,\cdot)$ is given in \cite[\textit{Lemma 3}]{Ian}.
\begin{proof}
Please refer to Appendix B.
\end{proof}

A lower bound on the achievable ergodic secrecy rate of the $k^{\rm th}$ MT in the local cell for the $\mathcal{N}$/$\mathcal{R}$ methods is obtained by combining (\ref{eq8}), (\ref{gammadl}), and (\ref{Cint}). However, the expression for the
ergodic capacity of the eavesdropper in (\ref{Cint}) is somewhat cumbersome and offers little insight into the impact of the various system parameters. Hence, in the next subsection, we derive a simple and tight upper bound for $C^{\rm eve}_{nk}$.
\subsection{Tight Upper Bound on the Ergodic Capacity of the Eavesdropper\label{s2f}}
In the following theorem, we provide a tight upper bound for the ergodic capacity of the eavesdropper.

\textit{Theorem 2}: For $N_t \to \infty$ and both the $\mathcal{N}$ and the $\mathcal{R}$ AN shaping matrix  generation methods, the ergodic capacity of the eavesdropper in (\ref{Ceve}) is upper bounded
by \footnote{We note that, strictly speaking, we have not proved that (\ref{Cup}) is a bound since we used an approximation for its derivation, see Appendix C. However, this approximation
is known to be very accurate \cite{appro} and comparisons of (\ref{Cup}) with simulation results for various system parameters suggest that (\ref{Cup}) is indeed an upper bound.}
\begin{equation}
\label{Cup}
 C_{nk}^{\rm eve} < \overline{C}_{nk}^{\rm eve} \approx \log_2 \left(1+\frac{\alpha}{\eta a(1-\beta)-c \eta \alpha/a}\right)=\log_2\left(\frac{(1-\zeta)\phi+\zeta}{-\zeta \phi+\zeta}\right),
\end{equation}
if $\beta < 1-c \alpha/a^2$, where we introduce the definitions $a=1+\rho(M-1)$, $c=1+\rho^2(M-1)$, and $\zeta=\frac{a \beta}{\alpha}-\frac{\beta c}{a(1-\beta)}$.
\begin{proof}
Please refer to Appendix C.
\end{proof}

\textit{Remark 1:} We note that a finite eavesdropper capacity results only if matrix ${\bf X}$ in (\ref{Ceve}) is invertible. Since ${\bf H}^{\rm eve}_{m}$, $m=1,\ldots,M$, are independent matrices with i.i.d.~entries, ${\bf X}$
is invertible if $M(N_t-K)\le N_e$ or equivalently $\beta\le1-\alpha/M$. Regardless of the values of $M$ and $\rho$, we have
\begin{equation}
\label{cond}
1-\alpha/[1+\rho^2(M-1)]\leq 1-c \alpha/a^2 \leq 1-\alpha/M.
\end{equation}
For $M=1$ or $\rho=1$, equality holds in (\ref{cond}). For $M>1$ and $\rho<1$, the condition for $\beta$ in Theorem 2 is in general stricter than the invertibility condition for ${\bf X}$.
Nevertheless, the typical operating region for a massive MIMO system is $\beta\ll 1$ \cite{survey, noncooperative}, where the upper bound in Theorem 2 is applicable.

Eq.~(\ref{Cup}) reveals that $\overline{C}_{nk}^{\rm eve}$ is monotonically increasing in $\alpha$, i.e., as expected, the eavesdropper can enhance his eavesdropping capability by deploying more antennas. Furthermore, in the
relevant parameter range, $0<\beta < 1-c \alpha/a^2$, $\overline{C}_{nk}^{\rm eve}$ is not monotonic in $\beta$ but a decreasing function for $\beta \in (0,1-\sqrt{c \alpha}/a)$ and an increasing function for $\beta \in (1-\sqrt{
c \alpha}/a,1-c \alpha/a^2)$. Hence, $\overline{C}_{nk}^{\rm eve}$ has a minimum at $\beta = 1-\sqrt{ c \alpha}/a$. Assuming $N_t$ and $N_e$ are fixed, this behaviour  can be explained as follows. For small $K$ (corresponding
to small $\beta$), the capacity of the eavesdropper is large because the amount of power allocated to the intercepted MT, $\phi P/K$, is large. As $K$ increases, the power allocated to the MT decreases which leads to a decrease
in the capacity. However, if $K$ is increased beyond a certain point, ${\bf X}$ becomes increasingly ill-conditioned which leads to an increase in the eavesdropper capacity.

Combining now (\ref{eq8}), (\ref{gammadl}), and (\ref{Cup}) gives a tight lower bound on the ergodic secrecy rate of the $k^{\rm th}$ MT in the local cell for both the $\mathcal{N}$ and the $\mathcal{R}$ methods. To gain more insight,
in the next two sections, we specialize the tight lower bound on the ergodic secrecy rate to the cases of perfect training and pilot contamination, respectively. This will allow us to further simplify the SINR expression of the $k^{\rm th}$ MT in the
local cell and the resulting ergodic secrecy rate expression.
\section{Performance Analysis for Perfect Training\label{s3}}
In this section, we analyze the secrecy performance of the considered downlink multi-cell massive MIMO system under the assumption of perfect CSI, i.e., $\hat{\bf h}_{nk} =\tilde{\bf h}_{nk}$, $k=1,\ldots,K$. To this end, for both considered
AN generation methods, we first simplify the lower bound on the achievable ergodic rate expression derived in Section \ref{s2d} by taking into account the perfect CSI assumption. Subsequently, exploiting this result, we derive simple and
insightful lower bounds on the achievable ergodic secrecy rate. Finally, we obtain an upper bound on the secrecy outage probability.
\subsection{Lower Bound on the Achievable Ergodic Rate\label{s3a}}
We first characterize some of the terms in (\ref{gammadl}) for the case of perfect training in the following lemma.

\textit{Lemma 2:} The received signal and interference powers at the $k^{\rm th}$ MT in the local cell can be expressed as
\begin{equation}
\mathbb{E}[\tilde{\bf h}_{nk} {\bf w}_{nk}]^2=\mathbb{E}^2[x] \quad {\rm and}\quad \mathbb{E}[|\tilde{\bf h}_{nk} {\bf w}_{mk}|^2]=\mathbb{E}[|\tilde{\bf h}_{nk} {\bf v}_{mi}|^2]=\mathbb{E}[y^2],\,\forall n\ne m
\end{equation}
respectively, where $x^2=\sum_{l=1}^{N_t} |u_l|^2\sim \chi^2_{2N_t}$, $y^2=|u_l|^2\sim \chi^2_{2}$, ${u_l}$ are i.i.d.~complex Gaussian r.v.s with zero mean and unit variance, and $\mathbb{E}[y^2]=1$.
\begin{proof}
Since each element of $\tilde{\bf h}_{nk}$ follows a Gaussian distribution with zero mean and unit variance and ${\bf w}_{nk}=\frac{{\bf h}_{nk}^H}{\|{\bf h}_{nk}\|} = \frac{\tilde{\bf h}_{nk}^H}{\|\tilde{\bf h}_{nk}\|}$,
$|\tilde{\bf h}_{nk} {\bf w}_{nk}|^2$ is a (scaled) chi-square r.v.~with $2N_t$ degrees of freedom and statistically equivalent to $x^2$. On the other hand, since ${\bf w}_{ml}$, $\forall \{m,l\} \neq \{n,k\}$, and ${\bf v}_{mi}$
are unit-norm vectors and independent of the small-scale fading vector $\tilde{\bf h}_{nk}$, the normalized interference terms, $|\tilde{\bf h}_{nk} {\bf w}_{mk}|^2$ and $|\tilde{\bf h}_{nk} {\bf v}_{mi}|^2$, are (scaled)
chi-square r.v.s with $2$ degrees of freedom and statistically equivalent to $y^2$.
\end{proof}

Introducing $x$ and $y$ in (\ref{gammadl}) and dividing both numerator and denominator by $p$, we obtain the SINRs for the $\mathcal{N}$ and $\mathcal{R}$ AN shaping matrices as
\begin{equation}
\label{gnk}
\gamma^{\mathcal{N}}_{nk}=\frac{l_{nk} \mathbb{E}^2[x]}{l_{nk}{\rm var}[x]+\eta \sum_{m \neq n}^M l_{mk} \sum_{i=1}^{N_t-K} \mathbb{E}[y^2]+\sum_{\{m,l\} \neq \{n,k\}} l_{mk} \mathbb{E}[y^2]+\frac{K}{\phi P}}
\end{equation}
and
\begin{equation}
\label{gnk3}
\gamma^{\mathcal{R}}_{nk}=\frac{l_{nk} \mathbb{E}^2[x]}{l_{nk}{\rm var}[x]+\eta \sum_{m=1}^M l_{mk} \sum_{i=1}^{N_t-K} \mathbb{E}[y^2]+\sum_{\{m,l\} \neq \{n,k\}} l_{mk} \mathbb{E}[y^2]+\frac{K}{\phi P}},
\end{equation}
respectively. The right hand sides of (\ref{gnk}) and (\ref{gnk3}) differ only in the second term of the denominator, where $\gamma^{\mathcal{R}}_{nk}$ contains an additional term $\eta l_{nk}\sum_{i=1}^{N_t-K}
\mathbb{E}[y^2]$, which is due to the AN leakage caused in the local cell. This term is absent in $\gamma^{\mathcal{N}}_{nk}$ as, for perfect CSI, the $\mathcal{N}$-method avoids AN leakage in the local cell.
Hence, $\gamma^{\mathcal{N}}_{nk} > \gamma^{\mathcal{R}}_{nk}$ always holds. Since for large $N_t$ we have \cite{pilotcontam}
\begin{equation}
\lim_{N_t \to \infty} \frac{\mathbb{E}^2[x]}{N_t}=1~{\rm and}~\lim_{N_t \rightarrow \infty} \frac{{\rm var}[x]}{N_t}=0,
\label{eq11}
\end{equation}
we obtain from (\ref{gnk}) and (\ref{gnk3})
\begin{equation}
\label{gnk2}
\lim_{N_t \rightarrow \infty}\gamma^{\mathcal{N}}_{nk}=\frac{l_{nk} N_t}{\eta \sum_{m \neq n}^M l_{mk}(N_t-K)+\sum_{\{m,l\} \neq \{n,k\}} l_{mk}+\frac{K}{\phi P}}
\end{equation}
and
\begin{equation}
\label{gnk4}
\lim_{N_t \rightarrow \infty} \gamma^{\mathcal{R}}_{nk}=\frac{l_{nk} N_t}{\eta \sum_{m=1}^M l_{mk}(N_t-K)+\sum_{\{m,l\} \neq \{n,k\}} l_{mk}+\frac{K}{\phi P}},
\end{equation}
respectively. In order to obtain simple yet insightful results, we adopt in the following a simplified path-loss model \cite{uldl}, similar to the simplified model introduced for the eavesdropper in Section \ref{s2e}. In particular,
we model the path-losses as $l_{mk}=1$ if $n=m$ and $l_{mk}=\rho$ if $n\ne m$, i.e., the path-loss between the local BS and the MTs in the local cell is $1$ and the path-loss between the BSs of the other cells and the MTs
in the local cell is  $\rho$. Hence, (\ref{gnk2}) and (\ref{gnk4}) simplify to
\begin{equation}
\label{bbb1}
\lim_{N_t \rightarrow \infty} \gamma^{\mathcal{N}}_{nk}=\frac{1}{(M-1)\rho(1-\beta)\eta+(M-1)\beta\rho+\beta+\frac{\beta}{\phi P}}
\end{equation}
and
\begin{equation}
\label{bbb2}
\lim_{N_t \rightarrow \infty} \gamma^{\mathcal{R}}_{nk}=\frac{1}{((M-1)\rho+1)(1-\beta)\eta+(M-1)\beta\rho+\beta+\frac{\beta}{\phi P}},
\end{equation}
respectively. The ergodic rate for the two considered AN shaping matrix generation methods is lower bounded by $\underline{R}_{nk}^{\Psi}=\log_2(1+\gamma^{\Psi}_{nk})$, where $\Psi \in \{\mathcal{N},\mathcal{R}\}$.
We note that for systems with few users, i.e., $\beta\to 0$, and $N_t\to\infty$, the lower bounds on the ergodic rate reduce to
\begin{equation}
\label{mtrate}
\underline{R}^{\mathcal{N}}_{nk} \approx \log_2 \left(1+\frac{1}{\eta(M-1)\rho}\right)\quad {\rm and}\quad \underline{R}^{\mathcal{R}}_{nk} \approx \log_2 \left(1+\frac{1}{\eta ((M-1)\rho+1)}\right),
\end{equation}
i.e., performance is limited by AN leakage. This is in contrast to massive MIMO systems without AN generation, whose performance in the considered regime ($\beta\to0$) is only limited by pilot contamination
\cite{survey, noncooperative},  which is not considered in this section but will be addressed in Section V. Moreover, (\ref{mtrate}) suggests that the performance difference between the $\mathcal{N}$-method
and the $\mathcal{R}$-method diminishes if the AN leakage from adjacent cells, which is proportional to $\eta(M-1)\rho$ for both methods, dominates the AN leakage for the $\mathcal{R}$-method in the
local cell, which is proportional to $\eta$.

Closed-form expressions for the lower bound on the achievable ergodic secrecy rate of the $k^{\rm th}$ MT in the local cell for the $\mathcal{N}$/$\mathcal{R}$ methods are obtained by combining (\ref{eq8}), (\ref{Cint}), and (\ref{bbb1})/(\ref{bbb2}).
The tightness of the proposed lower bounds will be confirmed in Section VI via simulations.
\subsection{Impact of System Parameters on Ergodic Secrecy Rate\label{s3b}}
In this subsection, we provide insight into the influence of the various system parameters on the ergodic secrecy rate. Combining (\ref{eq8}), (\ref{bbb1})/(\ref{bbb2}), and the upper bound on the ergodic secrecy
capacity in (\ref{Cup}), simple lower bounds for the ergodic secrecy rate valid for $N_t\to\infty$ are obtained as
\begin{eqnarray}
\label{low1}
\underline{R}_{nk}^{{\rm sec},\mathcal{N}}& =&\left[ \log_2\left(\frac{b \beta \zeta+(\beta+1-b \beta)\zeta \phi-(\beta+1)\zeta \phi^2}{b \beta \zeta+[\beta(1-\zeta)+b \beta \zeta]\phi+\beta (1-\zeta)\phi^2}\right)\right]^+,\\
\label{low2}
\underline{R}_{nk}^{{\rm sec},\mathcal{R}} &=&\left[ \log_2\left(\frac{(b+1)\beta \zeta+[1-(b+1)\beta]\zeta \phi-\zeta \phi^2}{(b+1)\beta \zeta+(b+1)\beta (1-\zeta)\phi}\right)\right]^+,
\end{eqnarray}
where $b=(M-1)\rho+1/P$ and $\eta=q/p=\beta(1/\phi-1)/(1-\beta)$ was used. In the following, we first investigate for what values of $\alpha$ a non-zero ergodic secrecy rate can be achieved.

\textbf{Impact of $\alpha$:} Let us denote the upper limit for $\alpha$ such that a positive secrecy rate can be achieved as $\alpha_{\rm sec}$. For the $\mathcal{N}$-method and the $\mathcal{R}$-method, we obtain
from (\ref{low1}) and (\ref{low2}), respectively, positive secrecy rates if $\alpha<\alpha_{\rm sec}^{\Psi}$, $\Psi\in \{{\cal N}, {\cal R}\}$, with
\begin{equation}
\label{alpha3}
\alpha_{\rm sec}^{\mathcal{N}} =\frac{a^2(1-\beta)}{ab(1-\beta)+c}\stackrel{\beta \to 0}{=}\frac{a}{b+c/a}=\frac{1+\rho(M-1)}{1/P+\rho(M-1)+c/a}
\end{equation}
and
\begin{equation}
\label{alpha1}
\alpha_{\rm sec}^{\mathcal{R}} = \frac{a^2(1-\beta)}{ a (b+1)(1-\beta)+c}\stackrel{\beta \to 0}{=}\frac{a}{b+1+c/a}=\frac{1}{1+1/[P(\rho(M-1)+1)]+c/a^2}.
\end{equation}
In both cases, $\alpha_{\rm sec}^{\Psi}$ is obtained for $\phi\to 0$, i.e., almost the entire transmit power is allocated to AN generation. For both methods, $\alpha_{\rm sec}$ is monotonically decreasing in $\beta$.
Furthermore, we always have $\alpha_{\rm sec}^{\mathcal{R}} <\alpha_{\rm sec}^{\mathcal{N}}$, i.e., the ${\cal N}$-method can tolerate a larger number of eavesdropper antennas than the ${\cal R}$-method at the
expense of a higher complexity in calculating the AN shaping matrix. The robustness of both AN shaping matrix designs can be improved by increasing the transmit power $P$. However, based on (\ref{alpha3}) and
(\ref{alpha1}) it can be shown that even for $P\to \infty$, the maximum values of $\alpha$ that yield a non-zero ergodic secrecy rate are limited as $\alpha_{\rm sec}^{\mathcal{N}} \le 4/3$ and $\alpha_{\rm sec}^{\mathcal{R}}\le 1$ regardless
of the choice of $M$ and $\rho$. We note that for a single-cell system with a single user, it was shown in \cite{khisti} that the ${\cal N}$-method can achieve non-zero secrecy rate for $\alpha< 2$. The smaller number of
tolerable eavesdropper antennas in the considered massive MIMO system are caused by the suboptimal matched-filter precoding at the base station, which was chosen for complexity reasons.

\textbf{Impact of $\phi$}: Eqs.~(\ref{low1}) and (\ref{low2}) reveal that zero secrecy rate results for $\phi=\phi_0=0$ and for a second value $\phi=\phi_{1}^\Psi$, $0<\phi_{1}^\Psi<1$, where
$\Psi\in \{{\cal N}, {\cal R}\}$. Specifically, $\phi_{1}^\Psi$ is given by
\begin{eqnarray}
\phi_{1}^{\mathcal{N}} &=&1- \frac{\alpha a(1-\beta)(b+1)}{a^2(1-\beta)(1+\alpha/a)-c \alpha}
\\
\phi_{1}^{\mathcal{R}} &=&1- \frac{\alpha a(1-\beta)(b+1)}{a^2(1-\beta)-c \alpha}
\end{eqnarray}
where $\phi_{1}^\Psi<1$ follows from the condition $\beta<1-c\alpha/a^2$ which is required for the validity of the upper bound on the ergodic secrecy capacity in (\ref{Cup}).
For $\phi=0$, all power is allocated to AN generation and no power is left for information transmission. On the other hand, for $\phi=\phi_{1}^\Psi$, the amount of AN generated is not
sufficient to prevent the eavesdropper from decoding the transmitted signal. This suggests that for $\alpha<\alpha_{\rm sec}^{\Psi}$,  $\Psi\in \{{\cal N}, {\cal R}\}$,
there exists an optimal $\phi$, $0< \phi< \phi_{1}^\Psi$, which maximizes the achievable ergodic secrecy rate. The values of the optimal $\phi$ can be obtained from (\ref{low1}) and (\ref{low2})
as
\begin{eqnarray}
\label{phistar}
\phi^{*}_{\mathcal{N}} &=&\frac{-(b\beta+b\zeta)+\sqrt{b(b+1)(\zeta-b \beta+\beta \zeta+b \beta \zeta)}}{1+b+\beta-b\zeta},
\\
\label{phistarR}
\phi^{*}_{\mathcal{R}}&=&\frac{-\zeta+\sqrt{\zeta-\beta-b\beta+\zeta \beta+b \beta \zeta}}{1-\zeta}.
\end{eqnarray}

\textbf{Impact of $\beta$:} It can be shown from (\ref{phistar}) and (\ref{phistarR}) that for both the $\mathcal{N}$ and $\mathcal{R}$ methods the optimal $\phi$ is a monotonically increasing function
of $\beta \in (0,1-c \alpha/a^2)$. Thus, as the number of MTs in the cell increase, the amount of power allocated to AN generation decreases. This can be explained by the fact that as $\beta$ increases,
the transmit power per MT used for information transmission, $\phi P/K$, decreases. To compensate for this effect, a larger $\phi$ is necessary. On the other hand, the ergodic secrecy rates for both the $
\mathcal{N}$ and $\mathcal{R}$ methods are decreasing functions of $\beta \in (0,1-c \alpha/a^2)$, cf.~(\ref{low1}), (\ref{low2}), i.e., as expected, for a given number of users the ergodic secrecy rates
increase with increasing number of BS antennas. Surprisingly, this property does not necessarily hold in case of pilot contamination, cf.~Section V.
\subsection{Secrecy Outage Probability Analysis\label{s3c}}
In delay limited scenarios, where one codeword spans only one channel realization, outages are unavoidable since Alice does not have the CSI of the eavesdropper channel and the secrecy outage probability has to be
used to characterize the performance of the system instead of the ergodic rate. For the considered multi-cell massive MIMO system, the rate of the desired user, $R_{nk}$, becomes deterministic as $N_t\to\infty$, but the
instantaneous capacity of the eavesdropper channel remains a random variable. A secrecy outage occurs whenever the target secrecy rate $R_{0}$ exceeds the actual instantaneous secrecy rate. Thus, the secrecy outage
probability of the $k^{\rm th}$ MT in the local cell is given by
\begin{equation}
\label{prob}
\varepsilon_{\rm out}={\rm Pr}\{R_{nk}-\log_2(1+\gamma_{\rm eve}) \leq R_{0}\}={\rm Pr}\{\gamma_{\rm eve} \geq 2^{R_{nk}-R_{0}}-1\}=1-F_{\gamma_{\rm eve}}(2^{R_{nk}-R_{0}}-1),
\end{equation}
where $\gamma_{{\rm eve}}=p{\bf w}_{nk}^H{\bf H}_n^{{\rm eve}H}{\bf X}^{-1}{\bf H}_n^{\rm eve}{\bf w}_{nk}$ and $F_{\gamma_{\rm eve}}(x)$ is given in Appendix B. A closed-form upper bound on the secrecy outage probability
is obtained by replacing $R_{nk}$ with $\underline{R}_{nk}^{\Psi}=\log_2(1+\gamma^{\Psi}_{nk})$ with $\gamma^{\Psi}_{nk}$ given in (\ref{bbb1})/(\ref{bbb2}).
\section{Performance Analysis for Pilot Contamination\label{s4}}
In this section, we analyze the performance of the considered multi-cell massive MIMO system for the case of pilot contamination. To this end, we simplify the lower bound on the achievable ergodic rate expression
derived in Section \ref{s2d} for the case of pilot contamination, derive insightful and tight lower bounds on the ergodic secrecy rate, and provide a closed-form expression for the secrecy outage probability.
\subsection{Lower Bound on the Achievable Ergodic Rate\label{s4a}}
The lower bound on the achievable ergodic rate of the users derived in Section \ref{s2d} is also applicable in case of pilot contamination. Thus, in a first step, we characterize the
four expectations/variances in the SINR expression in  (\ref{gammadl}).

Expressing the small-scale fading vector as $\tilde{\bf h}_{nk}=\hat{{\bf h}}_{nk}+{\bf e}_{nk}$, cf.~Section \ref{s2}, the denominator of (\ref{gammadl}) can be rewritten as
(we omit the path-loss for the moment)
\begin{equation}
\label{eq33}
\mathbb{E}[\tilde{\bf h}_{nk} {\bf w}_{nk}]=\mathbb{E}\bigg[\|\hat{{\bf h}}_{nk}\|+{\bf e}_{nk}\frac{\hat{{\bf h}}_{nk}}{\|\hat{{\bf h}}_{nk}\|}\bigg]=\mathbb{E}[\|\hat{{\bf h}}_{nk}\|]=\sqrt{\frac{p_\tau \tau l_{nk}}{1+p_\tau \tau \sum_{m=1}^M l_{mk}}}\mathbb{E}[x],
\end{equation}
where $x^2\sim \chi^2_{2N_t}$, cf.~Lemma 2. Furthermore, we observe from (\ref{est}) that, at the local BS, the channel estimate for the $k^{\rm th}$ MT in the local cell
involves the sum of all channel vectors between the local BS and the $k^{\rm th}$ MTs in all cells weighted with scaling factors $\frac{\sqrt{p_{\tau} \tau l_{mk}}}{1+p_{\tau} \tau \sum_{i=1}^M l_{ik}}$. Thus, the transmit beamforming
vector for the $k^{\rm th}$ MT in the local cell is also affected by the channel vectors between the local BS and the $k^{\rm th}$ MTs in all other cells. This is the fundamental problem introduced by pilot contamination. Using this
observation, the interference caused by the $k^{\rm th}$ MT in the $m^{\rm th}$ cell to the local cell (i.e., the component of the third term of the denominator in (\ref{gammadl}) with $l=k$) is given by
\begin{align}
\label{eq34}
& \hspace{0 cm} \nonumber\mathbb{E}[|\tilde{\bf h}_{mk} {\bf w}_{mk}|^2]=\mathbb{E}\big[\|\hat{{\bf h}}_{mk}\|^2\big]+\mathbb{E}\bigg[\frac{\hat{{\bf h}}_{mk}^H}{\|\hat{{\bf h}}_{mk}\|} {\bf e}_{mk}^H{\bf e}_{mk}\frac{\hat{{\bf h}}_{mk}}{\|\hat{{\bf h}}_{mk}\|}\bigg]\\
& \hspace{0 cm}\qquad =\frac{p_\tau \tau l_{mk}}{1+p_\tau \tau \sum_{p=1}^M l_{pk}}\mathbb{E}[x^2]+\frac{1+p_\tau \tau \sum_{p \neq m}l_{pk}}{1+p_\tau \tau \sum_{m=1}^M l_{mk}}\mathbb{E}[y^2],
\end{align}
where $y^2\sim \chi_2^2$, cf.~Lemma $1$. Exploiting now (\ref{eq33}) and (\ref{eq34}) and the definition of variance, i.e., ${\rm var}[x]=\mathbb{E}[x^2]-\mathbb{E}^2[x]$,
we obtain for the signal leakage term in (\ref{gammadl})
\begin{equation}
\label{eq35}
{\rm var}[\tilde{\bf h}_{nk} {\bf w}_{nk}] =\frac{p_\tau \tau l_{nk}}{1+p_\tau \tau \sum_{m=1}^M l_{mk}}{\rm var}[x]+\frac{1+p_\tau \tau \sum_{m \neq n}l_{mk}}{1+p_\tau \tau \sum_{m=1}^M l_{mk}}\mathbb{E}[y^2].
\end{equation}
Furthermore, the interference from the $l^{\rm th}$ MT, where $l\ne k$, in the adjacent (i.e., non-local) cells is given by
\begin{equation}
\label{eq36}
\mathbb{E}[|\tilde{\bf h}_{mk} {\bf w}_{ml}|^2]=\mathbb{E}[y^2],
\end{equation}
as each ${\bf w}_{ml}$, $\forall l \neq k$, has unit norm and is independent of  ${\bf h}_{mk}$. The inter-cell AN leakage is obtained as $\mathbb{E}[|\tilde{\bf h}_{mk} {\bf v}_{mi}|^2]=\mathbb{E}[y^2]$, $\forall m,i$,
as ${\bf v}_{mi}$ has unit norm and is independent of $\tilde{\bf h}_{mk}$. While the inter-cell AN leakage and the terms calculated in (\ref{eq33})-(\ref{eq36}) are identical for the $\mathcal{N}$ and $\mathcal{R}$
methods, the intra-cell AN leakage within the local cell depends on the AN shaping matrix design. In particular, for the $\mathcal{N}$-method, the AN is designed to lie in the null space of the estimated
channels of all $K$ MTs in the local cell. Thus, the intra-cell AN leakage is obtained as
\begin{equation}
\label{eq37}
\mathbb{E}[|\tilde{\bf h}_{nk} {\bf v}_{ni}|^2]=\mathbb{E}[{\bf v}_{ni}^H {\bf e}_{nk}^H {\bf e}_{nk} {\bf v}_{ni}]=\frac{1+p_\tau \tau \sum_{m \neq n}l_{mk}}{1+p_\tau \tau \sum_{m=1}^M l_{mk}} \mathbb{E}[y^2],
\end{equation}
due to the independence of ${\bf v}_{ni}$, $\forall i$, and ${\bf e}_{nk}$. On the other hand, for the $\mathcal{R}$-method, the AN is generated randomly, such that $\mathbb{E}[|\tilde{\bf h}_{nk} {\bf v}_{ni}|^2]=\mathbb{E}[y^2]$,
since the ${\bf v}_{ni}$, $\forall i$, have unit norm and are independent of $\tilde{\bf h}_{nk}$.

Plugging all intermediate results derived in this section so far into (\ref{gammadl}), we obtain
\begin{equation}
\label{eq38}
\gamma^{\mathcal{N}}_{nk}=\frac{\lambda_{nk}\mathbb{E}^2[x]}{\lambda_{nk}{\rm var}[x]+\sum_{m=1}^M\left(\mu_{mk}+\eta \sum_{i=1}^{N_t-K} \hat{\mu}_{mk}+\sum_{l \neq k} l_{mk}\right)\mathbb{E}[y^2]+\sum_{m \neq n}\lambda_{mk}\mathbb{E}[x^2]+\frac{K}{\phi P}}
\end{equation}
and
\begin{equation}
\label{eq39}
\gamma^{\mathcal{R}}_{nk}=\frac{\lambda_{nk}\mathbb{E}^2[x]}{\lambda_{nk}{\rm var}[x]+\sum_{m=1}^M\left(\mu_{mk}+\eta \sum_{i=1}^{N_t-K} l_{mk}+\sum_{l \neq k} l_{mk}\right)\mathbb{E}[y^2]+\sum_{m \neq n}\lambda_{mk}\mathbb{E}[x^2]+\frac{K}{\phi P}},
\end{equation}
where $\lambda_{mk}=\frac{p_\tau \tau l_{mk}^2}{1+p_\tau \tau \sum_{p=1}^M l_{pk}}$, $\mu_{mk}=l_{mk}\frac{1+p_\tau \tau \sum_{p \neq m}l_{pk}}{1+p_\tau \tau \sum_{m=1}^M l_{mk}}$, and
$\hat{\mu}_{mk}=\begin{cases}\mu_{mk},&m=n,\\l_{mk},&{\rm otherwise}\end{cases}$. Adopting now the same simplified interference model as in Section \ref{s3}, (\ref{eq38}) and (\ref{eq39}) can be
further simplified, and for large $N_t$, the corresponding lower bound on the achievable ergodic rates are given by
\begin{equation}
\label{rnull}
\underline{R}^{\rm \mathcal{N}}_{nk}=\log_2\left(1+\frac{\lambda}{(a-\lambda)(1-\beta)\eta+b\beta+(M-1)\rho^2\lambda+\frac{\beta}{\phi P}}\right)
\end{equation}
and
\begin{equation}
\label{rrnd}
\underline{R}^{\rm \mathcal{R}}_{nk}=\log_2\left(1+\frac{\lambda}{a(1-\beta)\eta+b\beta+(M-1)\rho^2\lambda+\frac{\beta}{\phi P}}\right),
\end{equation}
where $\lambda=\frac{p_\tau \tau}{1+p_\tau \tau a}$. From (\ref{rnull}) and (\ref{rrnd}) we observe that $\underline{R}^{\rm \mathcal{N}}_{nk}>\underline{R}^{\rm \mathcal{R}}_{nk}$ always holds but the
performance difference diminishes if $a\gg \lambda$. We note that for both AN shaping matrix designs the powers of the AN  leakage originating from other cells and
the inter-cell interference are proportional to $a-1=(M-1)\rho$. Furthermore, for the ${\cal N}$-method and the ${\cal R}$-method, the AN leakage originating in the local cell is
proportional to $(1-\lambda)\eta$ and $\eta$, respectively. Therefore, $a\gg \lambda$ implies that the AN leakage originating from other cells and the inter-cell interference are much
stronger than the AN leakage in the local cell and/or the pilot power $p_\tau$ is not sufficiently large to prevent AN leakage for the ${\cal N}$-method in the local cell. Furthermore, for $\beta\to0$,
we obtain $\underline{R}^{\rm \mathcal{N}}_{nk}=\log_2(1+\lambda/[(a-\lambda)\eta+(M-1)\rho^2\lambda])$ and $\underline{R}^{\rm \mathcal{R}}_{nk}=\log_2(1+\lambda/[a\eta+(M-1)\rho^2\lambda])$, i.e.,
in the asymptotic regime where the number of users is constant but the number of BS antennas increases without bound, the performance for both AN shaping matrix designs
is limited by both AN leakage and pilot contamination.

Since the ergodic capacity of the eavesdropper is not affected by the imperfect CSI at the local BS, a lower bound on the ergodic secrecy rate for pilot contamination can be calculated
from (\ref{eq8}), (\ref{Ceve}), and (\ref{rnull})/(\ref{rrnd}).
\subsection{Impact of System Parameters on Ergodic Secrecy Rate \label{s4b}}
To gain more insight, we employ again the upper bound on the ergodic capacity of the eavesdropper provided in Theorem 2. Combining (\ref{eq8}), (\ref{Cup}),
(\ref{rnull}), and (\ref{rrnd}), we obtain simple lower bounds for the ergodic secrecy rate for the $\mathcal{N}$ and $\mathcal{R}$ methods as $\underline{R}^{{\rm sec},{\rm \mathcal{N}}}_{nk}=$
\begin{equation}
\label{secnull1}
\left[\log_2\left(\frac{(b+1-\lambda)\beta \zeta+[(\beta+c)\lambda-(b+1-\lambda)\beta]\zeta \phi-\zeta(\beta+c)\lambda \phi^2}{(b+1-\lambda)\beta \zeta +[(\beta+c-1)\lambda \zeta+(b+1-\lambda)(1-\zeta)]\phi+(1-\zeta)(\beta+c-1)\lambda \phi^2}\right)\right]^+,
\end{equation}
and
\begin{equation}
\label{secrnd1}
\underline{R}^{{\rm sec},{\rm \mathcal{R}}}_{nk}=\left[\log_2\left(\frac{(b+1)\beta \zeta+[c\lambda-(b+1)\beta]\zeta \phi-\zeta c \lambda \phi^2}{(b+1)\beta \zeta +[(c-1)\lambda \zeta+(b+1)(1-\zeta)]\phi+(1-\zeta)(c-1)\lambda \phi^2}\right)\right]^+,
\end{equation}
respectively.

In the following, we investigate the impact of the system parameters on the ergodic secrecy rate in detail.

\textbf{Impact of $\alpha$:} Similar to the perfect training case we investigate in the following the upper limit for $\alpha$ such that a positive secrecy rate can be achieved.
We observe from (\ref{secnull1}) and (\ref{secrnd1}) that  a non-zero secrecy rate can be achieved as long as $\alpha<\alpha_{{\rm sec}}^\Psi$ holds where
\begin{eqnarray}
\label{eq46}
\alpha_{{\rm sec}}^{\mathcal{N}}&=& \frac{ a^2 (1-\beta)\lambda}{ a (1-\beta)(1+b-\lambda)+c \lambda} \stackrel{\beta \to 0}{=} \frac{ a^2 \lambda}{ a (1+b-\lambda)+c \lambda},
\\
\label{eq46a}
\alpha_{{\rm sec}}^{\mathcal{R}}&=& \frac{ a^2 (1-\beta) \lambda}{a(1-\beta)(1+b)+c \lambda}\stackrel{\beta \to 0}{=} \frac{ a^2 \lambda}{a(1+b)+c \lambda}.
\end{eqnarray}
Eqs.~(\ref{eq46}) and (\ref{eq46a}) reveal that the robustness of the considered multi-cell MIMO system to eavesdropping is monotonically decreasing with increasing number
of MTs in the system. On the other hand, allocating more resources to training, i.e., increasing $\lambda$ by increasing the pilot power, $p_\tau$, or the pilot sequence duration, $\tau$, leads to a higher
robustness against eavesdropping, i.e., a larger number of eavesdropper antennas can be tolerated. Furthermore, as expected, $\alpha_{{\rm sec}}^{\mathcal{N}}>\alpha_{{\rm sec}}^{\mathcal{R}}$,
i.e., the more complex $\mathcal{N}$-method is more robust to eavesdropping than the simple $\mathcal{R}$ method. However, $\alpha_{{\rm sec}}^{\mathcal{R}}$ approaches $\alpha_{{\rm sec}}^{\mathcal{N}}$
if $\lambda$ is small, i.e., both methods have a similar robustness to eavesdropping in case of strong pilot contamination since, in this case, the $\mathcal{N}$-method can no longer avoid
AN leakage within the local cell. We also note that, as expected, since $\lambda\le 1/b\le 1$ always holds, for a given AN shaping matrix design, the maximum tolerable number of eavesdropper antennas in case of
pilot contamination is always smaller than that in case of perfect training, cf.~(\ref{alpha1}), (\ref{alpha3}), and (\ref{eq46}), (\ref{eq46a}).

\textbf{Impact of $\phi$:} Similar to the case of perfect training, the ergodic secrecy rate for both AN shaping matrix designs becomes zero for $\phi=\phi_0= 0$ also for the case of pilot contamination,
cf.~(\ref{secnull1}) and (\ref{secrnd1}), since zero power is allocated to information transmission in this case. A second zero of the ergodic secrecy rate occurs for $\phi=\phi_{1}^\Psi$, $0<\phi_{1}^\Psi<1$, where
$\Psi\in \{{\cal N}, {\cal R}\}$. $\phi_{1}^\Psi$ is obtained from (\ref{secnull1}) and (\ref{secrnd1}) as
\begin{eqnarray}
\phi_{1}^{\cal N}&=& 1-\frac{\alpha a(\beta-1)((b+1)\beta+\lambda(c-1))}{\lambda(a(a+\alpha)\beta^2+(-a^2+\alpha(c-2)a+c\alpha)\beta-a\alpha(c-1))} \\
\phi_{1}^{\cal R}&=&1-\frac{\alpha a (\beta-1) ((b+1)\beta+\lambda(c-1))}{\lambda(a^2\beta^2+(-a^2+a\alpha(c-1)+c\alpha)\beta-a\alpha(c-1))}.
\end{eqnarray}
Furthermore, assuming $\alpha<\alpha_{\rm sec}^{\Psi}$ and taking the derivatives of (\ref{secnull1}) and (\ref{secrnd1}) with respect to $\phi$ and setting them to zero,
we obtain the optimal power allocation factors for the $\mathcal{N}$ and $\mathcal{R}$ methods as
\begin{align}
\label{eq43}
& \hspace{0 cm} \nonumber \phi_{\mathcal{N}}^*=\frac{-\sqrt{(b+1-\lambda)((-1+c)\lambda+(b+1)\beta)\beta((\beta+c\zeta)\lambda+(-1+\zeta)\beta(b+1))\lambda}}{(-\lambda\beta^2+((2-2c-\zeta)\lambda+(-1+\zeta)(b+1))\beta-c\lambda(-1+c))\lambda}\\
& \hspace{0 cm}+\frac{(-\lambda^2+(b+1)\lambda)\beta^2+((-c-\zeta+1)\lambda^2+((\zeta-1+c)b+\zeta-1+c)\lambda)\beta}{(-\lambda\beta^2+((2-2c-\zeta)\lambda+(-1+\zeta)(b+1))\beta-c\lambda(-1+c))\lambda}
\end{align}
and
\begin{equation}
\label{eq44}
\phi_{\mathcal{R}}^*=\frac{-\sqrt{\lambda((-1+c)\lambda+(b+1)\beta)(b+1)(c \zeta \lambda+(-1+\zeta)\beta(b+1))\beta}+((\zeta-1+c)b+\zeta-1+c)\lambda\beta}{\lambda((-1+\zeta)\beta(b+1)-c\lambda(-1+c))}.
\end{equation}

\textbf{Impact of $\beta$:} Based on (\ref{eq43}) and (\ref{eq44}) it can be shown that, similar to the case for perfect training, for pilot contamination, the optimal $\phi_{\mathcal{N}}^*$ and  $\phi_{\mathcal{R}}^*$ are
monotonically increasing in $\beta$. Furthermore, in Section \ref{s3}, we found that, for perfect training, the ergodic secrecy rate is monotonically increasing for decreasing $\beta$. However, for a given $\phi$, it can
be shown based on (\ref{secnull1}) and (\ref{secrnd1}) that this is no longer true in case of pilot contamination. In other words, if $\phi$ and the number of users $K$ are fixed, in case of pilot contamination, the ergodic
secrecy rate is not maximized by making the number of BS antennas, $N_t$, exceedingly large (i.e., $N_t\gg K$ such that $\beta \to 0$). Instead, there is an optimal finite number of BS antennas. We will investigate this
issue numerically in Section VI.

\textbf{Impact of $\lambda$:} Pilot contamination impacts the ergodic secrecy rate via $\lambda$, where smaller values of $\lambda$ imply that the MTs expend less resources for uplink training (i.e., they employ a smaller pilot
power $p_\tau$ and/or a shorter pilot sequence length, $\tau$). First, we observe from (\ref{secnull1}) and (\ref{secrnd1}) that both $\underline{R}^{{\rm sec},{\rm \mathcal{N}}}_{nk}$ and $\underline{R}^{{\rm sec},{\rm \mathcal{R}}}_{nk}$
are increasing functions of $\lambda$, i.e., as expected, if the MTs employ a higher pilot power and/or a longer pilot sequence for channel estimation, the ergodic secrecy rate improves. Furthermore, $\alpha_{\rm sec}$ is an increasing function
of $\lambda$, i.e., a higher uplink training power and/or longer pilot sequence lengths increase the operating region of the system where a non-zero secrecy rate can be achieved.

On the other hand, for a given coherence interval $T$, fixed transmit power $P$, and fixed pilot power $p_\tau$, the fraction of time allocated for training $\tau/T$ (and as a consequence $\lambda$) can be optimized for maximization of the net
ergodic secrecy rate given by $(1-\tau/T)R^{{\rm sec},{\Psi}}_{nk}$, $\Psi\in \{{\cal N}, {\cal R}\}$. We assume that the channels are constant within one coherence interval but change from one coherence interval to the next.
We also emphasize that by using the (net) ergodic secrecy rate as a performance measure, we implicitly assume coding over many coherence intervals. For small $\tau$, the factor $(1-\tau/T)$ is large but the ergodic secrecy
rate, $R^{{\rm sec},{\Psi}}_{nk}$, is small because of the unreliable channel estimation. On the other hand, for large $\tau$,  the factor $(1-\tau/T)$ is small but the ergodic secrecy rate, $R^{{\rm sec},{\Psi}}_{nk}$, is large
because of the more accurate channel estimation. Hence, $\tau$ can be optimized for optimal performance \cite{ud}. The optimization of $\tau$ will be studied numerically in Fig.~\ref{Fig8} in Section \ref{s5}.
\subsection{Secrecy Outage Probability Analysis\label{s4c}}
Plugging (\ref{rnull}) and (\ref{rrnd}) into the secrecy outage probability expression derived in (\ref{prob}), we obtain an upper bound for the secrecy outage probability for the case of pilot contamination as
\begin{equation}
\label{outagepc}
\overline{\varepsilon}^{\Psi}_{\rm out}=1-F_{\gamma_{\rm eve}}(2^{\underline{R}^{\Psi}_{nk}-R_{0}}-1),
\end{equation}
where $\Psi\in \{\mathcal{N},\mathcal{R}\}$.
\section{Numerical Examples\label{s5}}
In this section, we evaluate the secrecy performance of the considered multi-cell massive MIMO systems based on the analytical expressions derived in Sections III-V and via Monte-Carlo simulation.
We consider a system with $M=7$ hexagonal cells and adopt the simplified path-loss model, i.e., the severeness of the inter-cell interference is characterized by parameter $\rho$ only. The Monte-Carlo
simulation results for the ergodic secrecy rate of the $k^{\rm th}$ MT in the local cell  are based on (\ref{eq8}) where the achievable ergodic rate $R_{nk}$ is obtained from (\ref{rate}) and the ergodic secrecy capacity of the
eavesdropper is obtained from (\ref{Ceve}). Thereby, the expected values in (\ref{rate}) and (\ref{Ceve}) were evaluated by averaging over $3000$ random channel realizations. The Monte-Carlo simulation results for the outage
probability are obtained from $\varepsilon_{\rm out}={\rm Pr}\{R_{nk}-\log_2(1+\gamma_{\rm eve}) \leq R_{0}\}$, which was evaluated again based on $3000$ random channel realizations. The values of all relevant system parameters
are provided in the captions of the figures.
\subsection{Ergodic Secrecy Rate and Secrecy Outage Probability}
For the results shown in this section, we adopt a fixed power allocation factor of $\phi=0.75$. The optimization of $\phi$ will be addressed in the next subsection.

\begin{figure}
  \centering
    \includegraphics[width=4in]{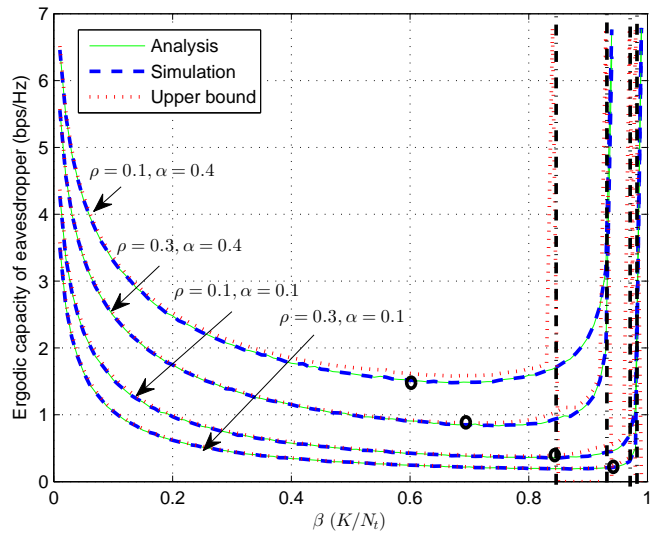}\\
    \caption{Ergodic capacity of the eavesdropper seeking to decode the information intended for the $k^{\rm th}$ MT in the local cell vs. the normalized number of MTs in the cell, $\beta$, for a system with total transmit power
    $P= 10$ dB, $M=7$, $\phi=0.75$, and $N_t=100$.}\label{Fig0}
  \end{figure}

In Fig.~\ref{Fig0}, we verify the derived analytical expressions for the ergodic capacity of the eavesdropper which seeks to decode the information intended for the $k^{\rm th}$ MT in the local cell. The analytical results were generated with
(\ref{Cint}) while the upper bound results were computed with (\ref{Cup}). The vertical dashed lines denote $\beta=1-c\alpha/a^2$. Fig.~\ref{Fig0} reveals that for $\beta<1-c \alpha/a^2$, the upper bound is very tight.
For $1-c \alpha/a^2<\beta<1-\alpha/M$, the upper bound is not applicable,
although the ergodic capacity of the eavesdropper is still finite, cf.~Theorem 2 and Remark 1. For $\beta\to 1-\alpha/M$, the ergodic capacity of the eavesdropper tends to infinity since ${\bf X}$ becomes singular. Furthermore,  we
observe from Fig.~\ref{Fig0} that increasing inter-cell interference (i.e., larger inter-cell interference factors, $\rho$) has a negative effect on the ergodic capacity of the eavesdropper, whereas as expected, the eavesdropper can improve
his performance by adding more antennas, $N_e$ (i.e., by increasing $\alpha$). Moreover, Fig.~\ref{Fig0} confirms that the ergodic capacity of the eavesdropper is monotonically decreasing in $\beta$ in the interval
$(0,1-\sqrt{c \alpha}/a)$ and monotonically increasing in $\beta$ in the interval $(1-\sqrt{c \alpha}/a,1-c\alpha/a^2)$. The resulting minimum of the ergodic capacity of the eavesdropper at $\beta=1-\sqrt{c \alpha}/a$ is
denoted by a black circle in Fig.~\ref{Fig0}.

\begin{figure}
  \centering
    \includegraphics[width=4in]{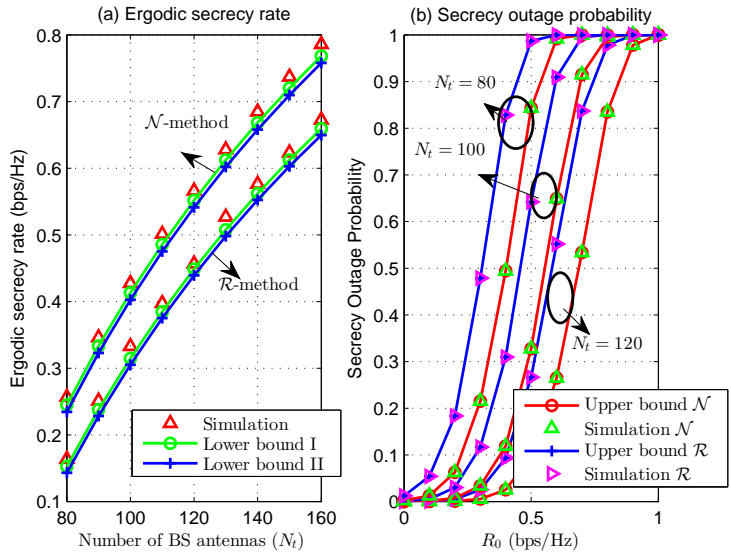}\\
    \caption{Ergodic secrecy rate and outage probability for perfect training, $M=7$, $P=10$ dB, $K=10$, $\rho=0.3$, $\alpha=0.1$, and $\phi=0.75$.}\label{Fig2}
  \end{figure}

In Fig.~\ref{Fig2}, for the case of perfect training, we show the ergodic secrecy rate vs.~the number of BS antennas (subfigure (a)) and the secrecy outage probability vs.~the target secrecy rate $R_0$ (subfigure (b)) for the $k^{\rm th}$ MT in the local cell.
Results for both considered AN shaping matrix designs are shown. In subfigure (a), lower bound I was obtained based on (\ref{eq8}), (\ref{Cint}), (\ref{bbb1}), and (\ref{bbb2}) and lower bound II was obtained with (\ref{low1}) and (\ref{low2}).
In subfigure (b), the upper bound was obtained with (\ref{prob}). Fig.~\ref{Fig2} reveals that the derived bounds for the ergodic secrecy rate and the secrecy outage probability are accurate. As expected, for the ergodic secrecy rate,
lower bound I is somewhat tighter than lower bound II. Furthermore, increasing the number of BS antennas $N_t$ improves both the ergodic secrecy rate as well as the secrecy outage probability. Moreover, as expected, the
$\mathcal{N}$-method for generation of the AN shaping matrix always outperforms the $\mathcal{R}$-method as the $\mathcal{N}$-method avoids intra-cell AN leakage.

\begin{figure}
  \centering
    \includegraphics[width=4in]{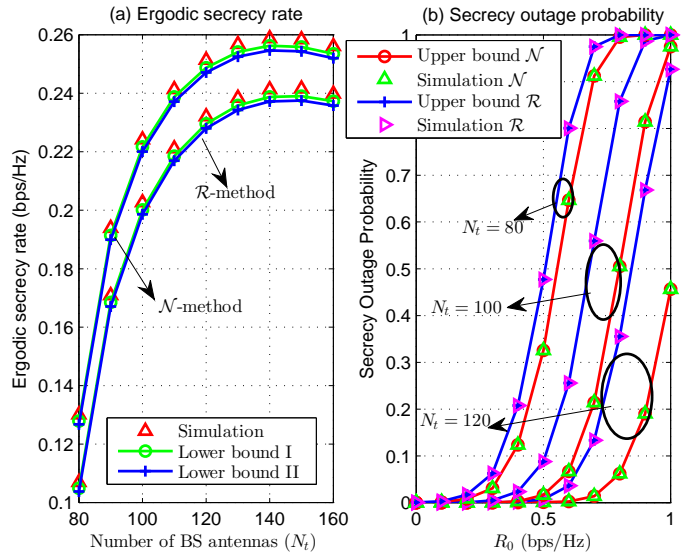}\\
    \caption{Ergodic secrecy rate and outage probability for pilot contamination, $M=7$, $P=10$ dB, $K=10$ MTs, $\rho=0.1$, $\alpha=0.1$, $\phi=0.75$, $\tau=K$, and $p_\tau=P/K$. }
   \label{Fig3}
\end{figure}

In Fig.~\ref{Fig3}, we show the same performance metrics as in Fig.~\ref{Fig2}, however, now for the case of pilot contamination. In subfigure (a), lower bound I was obtained based on (\ref{eq8}), (\ref{Cint}), (\ref{rnull}), and (\ref{rrnd}),
whereas lower bound II was obtained with (\ref{secnull1}) and (\ref{secrnd1}). In subfigure (b), the upper bound was obtained with (\ref{outagepc}). Similar to the case of perfect training, the derived bounds on the ergodic secrecy rate and the
secrecy outage probability are very tight. A comparison of Figs.~\ref{Fig2} and \ref{Fig3} reveals that pilot contamination causes a significant performance degradation in terms of both ergodic secrecy rate and secrecy outage probability.
Furthermore, unlike for the case of perfect training, for pilot contamination, the ergodic secrecy rate is not monotonically increasing in $N_t$ but has a unique maximum for both AN shaping matrix designs.
\subsection{Optimal Power Allocation}
In this subsection, we investigate the optimization of power allocation factor $\phi$ and illustrate its impact on the ergodic secrecy rate.

\begin{figure}
  \centering
    \includegraphics[width=4in]{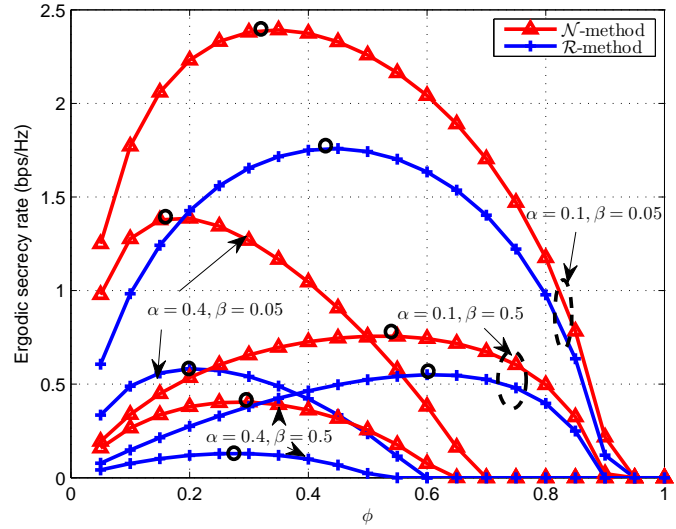}\\
    \caption{Ergodic secrecy rate vs.~power allocation factor $\phi$ assuming perfect training, $N_t=100$, $M=7$, $P=10$ dB, and $\rho=0.1$. Black circles denote the optimal power allocation factor, $\phi^\ast$, obtained with (\ref{phistar}) and (\ref{phistarR}).}\label{Fig4}
  \end{figure}

\begin{figure}
  \centering
    \includegraphics[width=4in]{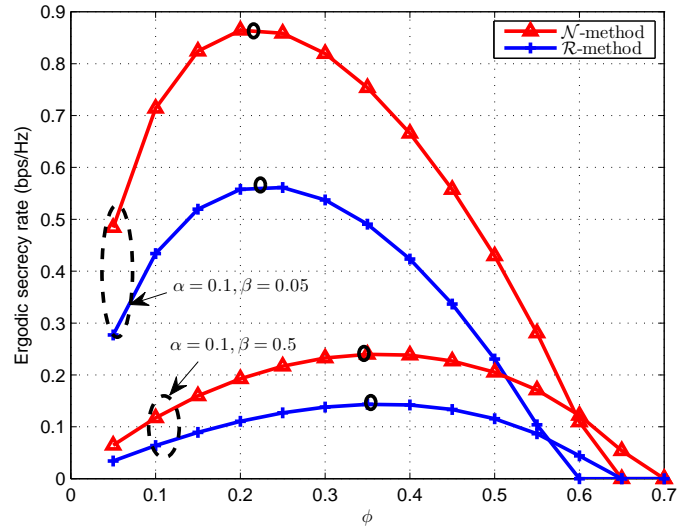}\\
    \caption{Ergodic secrecy rate vs.~power allocation factor $\phi$ assuming pilot contamination, $M=7$, $N_t=100$, $P=20$ dB, $\tau=K$, $p_\tau=P/K$, and $\rho=0.1$.
    Black circles denote the optimal power allocation factor, $\phi^\ast$, obtained with (\ref{eq43}) and (\ref{eq44}).}\label{Fig5}
  \end{figure}

Figs.~\ref{Fig4} and \ref{Fig5} show the ergodic secrecy rates of the $k^{\rm th}$ MT in the local cell as functions of $\phi$ for the cases of perfect training and pilot contamination, respectively. The ergodic secrecy rate curves were obtained via
Monte Carlo simulation and various values of $\alpha$ and $\beta$ are considered. The optimal values for $\phi$ obtained with (\ref{phistar})/(\ref{phistarR}) (for perfect training) and (\ref{eq43})/(\ref{eq44}) (for pilot contamination) are denoted by black circles. As expected from our discussions in Sections \ref{s3} and \ref{s4},  Figs.~\ref{Fig4} and \ref{Fig5} show that,  for both the ${\cal N}$ and the ${\cal R}$ AN shaping matrix desigs, the optimal $\phi^*$ is decreasing in
$\alpha$, i.e., the system should allocate more power to AN if the eavesdropper is becoming stronger, and increasing in $\beta$, i.e., less power should be allocated to AN if the number of users increases. For $\alpha=0.4$, no results are
shown for the case of pilot contamination in Fig.~\ref{Fig5} since the corresponding ergodic secrecy rates are zero for all choices of $\phi$, i.e., $\alpha> \alpha_{\sec}$ holds in this case.

\begin{figure}
\centering
    \includegraphics[width=4in]{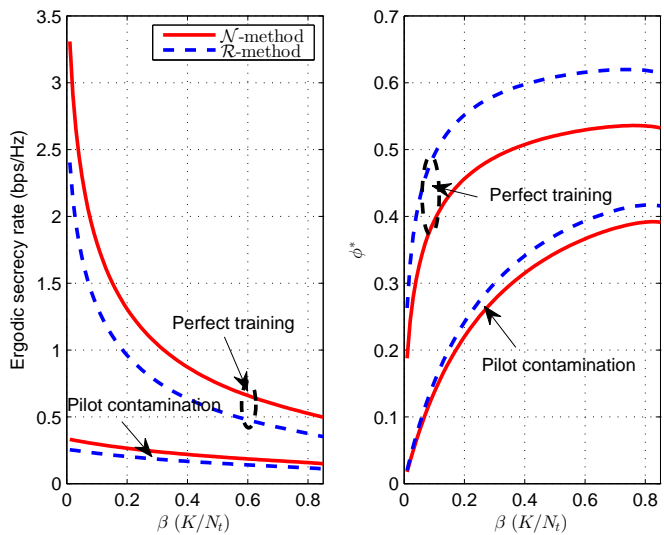}\\
    \caption{Ergodic secrecy rate and optimal power allocation factor, $\phi^*$, vs.~$\beta$ for $M=7$, $P=10$ dB, $N_t=100$, $\alpha=0.3$, and $\rho=0.1$. In case of pilot contamination,  $\tau=K$ and $p_\tau=P/K$. The ergodic secrecy rates
    were obtained with (\ref{low1}), (\ref{low2}), (\ref{secnull1}), and (\ref{secrnd1}). The optimal power allocation factor was obtained with (\ref{phistar}), (\ref{phistarR}), \ref{eq43}), and (\ref{eq44}).}\label{Fig6}
\end{figure}

In Fig.~\ref{Fig6}, we depict the ergodic secrecy rate and the optimal power allocation factor, $\phi^*$, as functions of the normalized number of MTs in each cell, $\beta$. Thereby, the ergodic secrecy rate is calculated using the optimal $\phi^*$,
which was obtained based on the analytical results in Sections IV and V for the case of perfect training and pilot contamination, respectively. We observe that, unlike the case when $\phi$ is fixed, if $\phi$ is optimized, the ergodic secrecy rate is
a non-increasing function of $\beta$ also in case of pilot contamination, i.e., for a given number of users, increasing the number of BS antennas is always beneficial. On the other hand, for all considered cases, the optimal value of $\phi$ is a
monotonically increasing function of $\beta$, i.e., as the number of users in the system increases relative to the number of BS antennas, less power is allocated to AN. Also, the performance gap between both AN shaping matrix design methods
decreases with increasing $\beta$.

\subsection{Conditions for Non-zero Ergodic Secrecy Rate}
\begin{figure}
  \centering
    \includegraphics[width=4in]{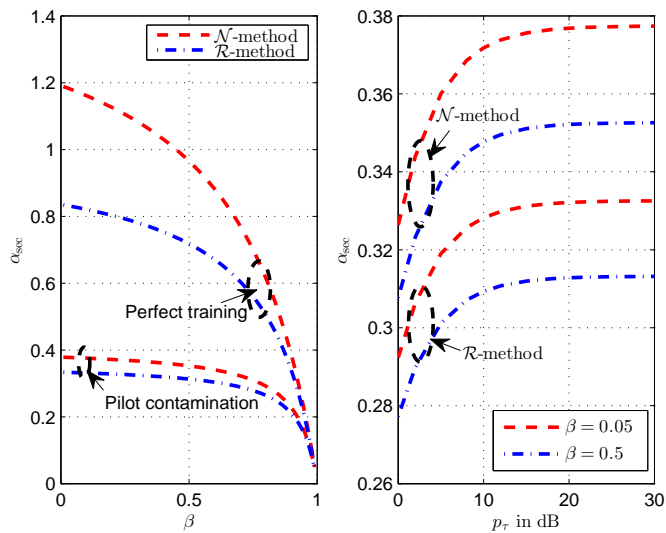}\\
    \caption{$\alpha_{\rm sec}$ vs.~$\beta$ and $p_{\tau}$ for pilot contamination, total transmit power $P=20$ dB, $M=7$, $N_t=100$, $\rho=0.1$, and $\tau=K$.}\label{Fig7}
  \end{figure}

In Fig.~\ref{Fig7}, we illustrate for both AN shaping matrix designs under what conditions a non-zero ergodic secrecy rate is possible. To this end, we plot $\alpha_{\rm sec}$ as defined in (\ref{alpha3}), (\ref{alpha1}), (\ref{eq46}), and
(\ref{eq46a}) as functions of $\beta$ for $p_\tau=P/K$ (subfigure on left hand side) and the amount of power, $p_\tau$, spent by the MTs for training for $\beta=0.05,0.5$ (subfigure on right hand side). For $\alpha\ge \alpha_{\rm sec}$, the ergodic
secrecy rate is zero regardless of the amount of power allocated to AN. On the other hand, for $\alpha < \alpha_{\rm sec}$, a positive ergodic secrecy rate can be achieved. We observe from Fig.~\ref{Fig7} that for both
AN shaping matrix designs $\alpha_{\rm sec}$ is a decreasing function of $\beta$, whereas it is an increasing function of $p_\tau$, i.e., the more reliable the channel estimates, the more eavesdropper antennas can be tolerated
before the ergodic secrecy rate drops to zero. However, $\alpha_{\rm sec}$ saturates for large values of $p_\tau$. We note that the values of $\alpha_{\sec}$ are smaller for the ${\cal R}$-method than for the ${\cal N}$-method
because of the larger intra-cell AN leakage caused by the ${\cal R}$-method.

\subsection{Optimization of the Net Ergodic Secrecy Rate}
\begin{figure}
  \centering
    \includegraphics[width=4in]{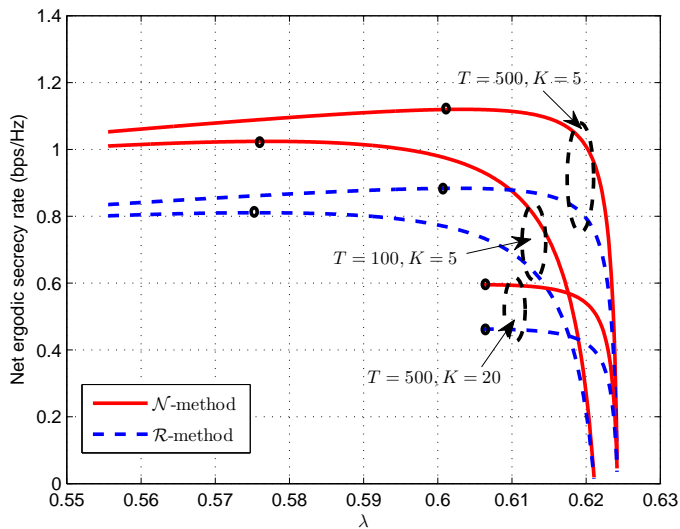}\\
    \caption{Net ergodic secrecy rate vs.~$\lambda$ for a system with optimal $\phi^*$, $N_t=100$, $M=7$, $\alpha=0.1$, $P=10$ dB, $p_\tau=0$ dB, and $\rho=0.1$. Black circles denote the maximum net ergodic secrecy rate.}\label{Fig8}
  \end{figure}

Fig.~\ref{Fig8} depicts the net ergodic secrecy rate, $(1-\tau/T)R^{{\rm sec}}_{nk}$, as a function of $\lambda$, where the lower bounds in (\ref{secnull1}) and (\ref{secrnd1}) were used to approximate $R^{{\rm sec}}_{nk}$. The cases of
$T=100$ and $T=500$ are considered for $K=5$ and $K=20$ MTs. We assume that $p_\tau = 0$ dB and $\lambda$ is varied by changing $\tau$ and the optimal power allocation factor $\phi^*$ is employed.
Thereby, the range of possible $\tau$ is $[K,T)$, which directly translates into the range of possible $\lambda$ as $\lambda=\frac{p_\tau \tau}{1+p_\tau \tau a}$. Fig.~\ref{Fig8} reveals that the optimal $\lambda$ is (slightly) increasing in $T$
since for larger values of $T$, more time for allocation to uplink training is available, i.e., $\tau$ can be increased resulting in a larger value for the optimal $\lambda$. For $K=20$, the lower limit of the permissible interval for $\tau$ given by 
$\tau=K$ yields the maximum net secrecy rate. In this case, increasing $\tau$ beyond $\tau=K$ does not improve $R^{{\rm sec}}_{nk}$ sufficiently to compensate for the decrease of the term $1-\tau/T$.

\section{Conclusions}
In this paper, we considered a multi-cell massive MIMO system with matched-filter precoding and AN generation at the BS for secure downlink transmission in the presence of a multi-antenna passive eavesdropper.
For AN generation, we considered both the conventional null space based AN shaping matrix design and a novel random AN shaping matrix design. For both perfect training and pilot contamination, we derived two tight
lower bounds on the ergodic secrecy rate and a tight upper bound on the secrecy outage probability. The analytical expressions allowed us to optimize the amount of power allocated to AN generation and to gain significant
insight into the impact of the system parameters on performance. In particular, our results reveal that for the considered multi-cell massive MIMO system with matched-filter precoding (1) AN generation is necessary to achieve
a non-zero ergodic secrecy rate if the user and the eavesdropper experience the same path-loss, (2) secrecy cannot be guaranteed if the eavesdropper has too many antennas, (3) for the case of pilot contamination, the
ergodic secrecy rate is only an increasing function of the number of BS antennas if the amount of power allocated to AN generation is optimized, and (4) the proposed random AN shaping matrix design is a promising
low-complexity alternative to the conventional null space based AN shaping matrix design.
\section*{Appendix}
\subsection{Proof of \textit{Lemma 1}\label{app-a}}
The proof closely follows \cite{GEYRC12}. We first derive an expression for the secrecy rate for given realizations of ${\bf h}_{mk}$ and ${\bf H}^{{\rm eve}}_m$, $k=1,\ldots, K$, $m=1,\ldots, M$. Since the MISOME channel
in (\ref{x1}) and (\ref{x2}) is a non-degraded broadcast channel \cite{khisti}, the secrecy capacity is given by \cite{GEYRC12}, \cite{CK78}
\begin{equation}
C_{nk}^{\rm sec}({\bf h})=\max_{s_{nk} \rightarrow {\bf w}_{nk}s_{nk}\rightarrow y_{nk},{\bf y}_{\rm eve}} I\left(s_{nk};y_{nk}|{\bf h}\right)-I\left(s_{nk};{\bf y}_{\rm eve}|{\bf h}\right),
\end{equation}
where vector ${\bf h}$ contains the CSI of all user and eavesdropper channels and $I(x;y|{\bf h})$ is the mutual information between two r.v.s $x$ and $y$ conditioned on the CSI vector. $C_{nk}^{\rm sec}({\bf h})$
is achieved by maximizing over all joint distributions such that a Markov chain $s_{nk} \rightarrow {\bf w}_{nk}s_{nk}\rightarrow y_{nk},{\bf y}_{\rm eve}$ results, where $s_{nk}$ is an arbitrary input variable \cite{GEYRC12}.
Specifically, for $s_{nk} \sim \mathbb{CN}(0,1)$ an achievable secrecy rate for the $k^{\rm th}$ MT in the local cell, $R_{nk}^{\rm sec} ({\bf h})$, is given by
\begin{eqnarray}
R_{nk}^{\rm sec}({\bf h})&=&\bigg[I\left(s_{nk};y_{nk}|{\bf h}\right)-I\left(s_{nk};{\bf y}_{\rm eve}|{\bf h}\right)\bigg]^+\stackrel{(a)}{=}\bigg[I\left({\bf w}_{nk}s_{nk};y_{nk}|{\bf h}\right)-I\left({\bf w}_{nk}s_{nk};{\bf y}_{\rm eve}|{\bf h}\right)\bigg]^+\nonumber
\\
&\stackrel{(b)}{\ge}& \bigg[R_{nk}\left({\bf h}\right)-C_{nk}^{\rm eve}\left({\bf h}\right)\bigg]^+
\end{eqnarray}
where (a) follows since ${\bf w}_{nk}s_{nk}$ is a deterministic function of $s_{nk}$. Furthermore, $R_{nk}({\bf h})\le \max I\left({\bf w}_{nk}s_{nk};y_{nk}|{\bf h}\right)$ is an achievable rate of the $k^{\rm th}$ MT in the local cell and
$C_{nk}^{\rm eve}({\bf h}) = \log_2 \left(1+p {\bf w}^H_{nk} {\bf H}^{{\rm eve}H}_{n} {\bf X}^{-1}{\bf H}^{\rm eve}_{n} {\bf w}_{nk}\right)\ge I\left({\bf w}_{nk}s_{nk};{\bf y}_{\rm eve}|{\bf h}\right)$ is an upper bound on the mutual
information $I\left({\bf w}_{nk}s_{nk};{\bf y}_{\rm eve}|{\bf h}\right)$. Thus, follows (b). We note that for computation of $C_{nk}^{\rm eve}({\bf h})$ we made the worst-case assumption that the eavesdropper can decode and cancel the signals of
all MTs except the signal intended for the MT of interest \cite[Chapter 10.2]{tse}.

Finally, to arrive at the ergodic secrecy rate, we average $R_{nk}^{\rm sec}({\bf h})$ over all channel realizations, which results in \cite{zhou}
\begin{eqnarray}
\label{eq8a}
\mathbb{E}\bigg[R_{nk}^{\rm sec}({\bf h})\bigg]&=&\mathbb{E}\left[ \bigg[R_{nk}\left({\bf h}\right)-C_{nk}^{\rm eve}\left({\bf h}\right)\bigg]^+\right]\nonumber
\\
&\ge& \bigg[\mathbb{E}\left[R_{nk}\left({\bf h}\right)\right]-\mathbb{E}\left[ C_{nk}^{\rm eve}\left({\bf h}\right)\right]\bigg]^+=R_{nk}^{\rm sec}.
\end{eqnarray}
Introducing the definitions of the achievable ergodic secrecy rate, $R_{nk}=\mathbb{E}\left[R_{nk}\left({\bf h}\right)\right]$, and the ergodic eavesdropper capacity, $C_{nk}^{\rm eve}=\mathbb{E}\left[ C_{nk}^{\rm eve}\left({\bf h}\right)\right]$, completes the proof.
\subsection{Proof of \textit{Theorem 1}\label{app-b}}
We first recall that the entries of ${\bf H}_m^{\rm eve}$, $m=1,\ldots,M$, are mutually independent complex Gaussian r.v.s. On the other hand, for $N_t\to \infty$ and both AN shaping matrix designs, the vectors ${\bf v}_{ml},l=1,\ldots,N_t-K$,
form an orthonormal basis. Hence, ${\bf H}_m^{\rm eve}{\bf V}_m$, $m=1,\ldots,M$, also has independent complex Gaussian entries, which are independent from the complex Gaussian entries of ${\bf H}_n^{\rm eve}{\bf w}_{nk}$. Thus,
the term $\gamma_{{\rm eve}}=p{\bf w}_{nk}^H{\bf H}_n^{{\rm eve}H}{\bf X}^{-1}{\bf H}_n^{\rm eve}{\bf w}_{nk}$ in (\ref{Ceve}) is equivalent to the SINR of an $N_e$-branch MMSE diversity combiner with $M(N_t-K)$ interferers \cite{gao,zhou}. As a result, for the considered simplified path-loss model, the cumulative density function (CDF) of the received SINR, $\gamma_{{\rm eve}}$, at the eavesdropper is given by \cite{gao}
\begin{equation}
\label{FXx}
F_{{\gamma_{\rm eve}}}(x)=\frac{\sum_{i=0}^{N_e-1} \lambda_i x^i}{\prod_{j=1}^{2} (1+\mu_j x)^{b_j}},
\end{equation}
where $\lambda_i$, $\mu_j$, and $b_j$ are defined in Theorem 1.  Exploiting (\ref{FXx}), we can rewrite (\ref{Ceve}) as
\begin{eqnarray}
\label{Cinta}
\hspace{0 cm} \nonumber C_{\rm eve}&\overset{(a)}{=}& \frac{1}{\ln 2} \int_0^{\infty} (1+x)^{-1} F_{{\gamma_{\rm eve}}}(x) dx\\
\hspace{0 cm} \nonumber &=& \frac{1}{\ln 2} \sum_{i=0}^{N_e-1} \lambda_i \times \int_0^{\infty} \frac{x^i}{(1+x)\prod_{j=1}^2 (1+\mu_j x)^{b_j}} dx\\
\hspace{0 cm} \nonumber &\overset{(b)}{=}& \frac{1}{\ln 2} \sum_{i=0}^{N_e-1} \lambda_i \times \frac{1}{\mu_0} \sum_{j=1}^2 \sum_{l=1}^{b_j} \int_0^{\infty}\frac{\omega_{jl}}{(x+1)(x+\frac{1}{\mu_j})^l}dx\\
\hspace{0 cm} &\overset{(c)}{=}& \frac{1}{\ln 2} \sum_{i=0}^{N_e-1} \lambda_i \times \frac{1}{\mu_0} \sum_{j=1}^2 \sum_{l=2}^{b_j} \omega_{jl}I(1/\mu_j,l),
\end{eqnarray}
where $\mu_0$, $\omega_{jl}$, and $I(\cdot,\cdot)$ are defined in Theorem 1. Here, (a) is obtained using integration by parts, (b) holds if the order of $x$ in the denominator of (\ref{FXx}) is not smaller than that in the numerator, i.e.,
$N_t-K\geq N_e/M$ or equivalently $1-\beta\geq \alpha/M$, which is also the condition to ensure invertibility of ${\bf X}$ in (\ref{Ceve}), and (c) is obtained using the definition of $I(\cdot,\cdot)$ given in Theorem 1.
This completes the proof.
\subsection{Proof of \textit{Theorem 2}}
Using Jensen's inequality and the mutual independence of $\tilde{\bf w}_{nk}={\bf H}_n^{\rm eve}{\bf w}_{nk}$ and ${\bf H}_m^{\rm eve}{\bf V}_m$, $m=1,\ldots, M$ (cf.~Appendix B), $C_{nk}^{\rm eve}$ in (\ref{Ceve})
is upper bounded by
\begin{equation}
\label{Cup2}
C_{nk}^{\rm eve} \leq \log_2 \left(1+\mathbb{E}_{\tilde{\bf w}_{nk}}\left[p \tilde{\bf w}_{nk}^H \mathbb{E}\left[{\bf X}^{-1}\right]\tilde{\bf w}_{nk}\right]\right).
\end{equation}
Let us first focus on the term $\mathbb{E}\left[{\bf X}^{-1}\right]$ in (\ref{Cup2}) and note that ${\bf X}$ is statistically equivalent to a weighted sum of two scaled Wishart matrices \cite{rmt}. Specifically, we have
${\bf X}=q {\bf X}_1 + \rho q {\bf X}_2$ with ${\bf X}_1 \sim \mathcal{W}_{N_e}(N_t-K,{\bf I}_{N_e})$ and ${\bf X}_2 \sim \mathcal{W}_{N_e}((M-1)(N_t-K),{\bf I}_{N_e})$, where $\mathcal{W}_{A}(B,{\bf I}_{A})$ denotes an
$A\times A$ Wishart matrix with $B$ degrees of freedom. Strictly speaking, ${\bf X}$ is not a Wishart matrix, and the exact distribution of ${\bf X}$ seems intractable. However, ${\bf X}$ may be accurately approximated
as a single scaled Wishart matrix, ${\bf X} \sim \mathcal{W}_{N_e}(\varphi,\xi {\bf I}_{N_e})$, where parameters $\xi$ and $\varphi$ are chosen such that the first two moments of ${\bf X}$ and $q {\bf X}_1 + \rho q {\bf X}_2$
are identical \cite{appro,wishart}. Equating the first two moments of the traces of these matrices yields \cite{wishart}
\begin{equation}
\label{eq47}
\xi \varphi=q(N_t-K)+\rho q (M-1)(N_t-K),
\end{equation}
and
\begin{equation}
\label{eq48}
\xi^2 \varphi=q^2(N_t-K)+\rho^2 q^2 (M-1)(N_t-K).
\end{equation}
By exploiting the expectation of an inverse Wishart matrix given in \cite[Eq.~(12)]{wishart}, we obtain $\mathbb{E}[{\bf X}^{-1}]=\frac{ 1}{\xi(\varphi-N_e-1)} {\bf I}_{N_e}$ with $\xi=cq/a$ if $\varphi -N_e>1$ or equivalently
if $\beta<1-c \alpha/a^2$ for $N_t \to \infty$. Plugging this result and $\mathbb{E}[\tilde{\bf w}^H_{nk}\tilde{\bf w}_{nk}]=N_e$ into (\ref{Cup2}), we finally obtain the result in (\ref{Cup}). This completes the proof.


\begin{thebibliography}{20}
\bibitem{sec_survey}
A. Mukherjee, S. A. A. Fakoorian, J. Huang, and A. L. Swindlehurst,
``Principles of physical-layer security in multiuser wireless networks: A
survey,'' submitted to \emph{IEEE Commun. Surveys and Tutorials}, 2013. arXiv:1011.3754 [cs.IT]

\bibitem{wiretap}
A. D. Wyner, ``The wire-tap channel,'' \emph{Bell Syst. Tech. J.}, vol. 54, no. 8, pp. 1355-1387, Oct. 1975.

\bibitem{khisti}
A. Khisti and G. Wornell, ``Secure transmission with multiple antennas
I: The MISOME wiretap channel,'' \emph{IEEE Trans. Inform. Theory}, vol. 56,
no. 7, pp. 3088-3104, Jul. 2010.

\bibitem{khisti2}
A. Khisti and G. Wornell, ``Secure transmission with multiple antennas II: The MIMOME wiretap channel,'' \emph{IEEE Trans. Inform. Theory}, vol. 56, no. 11, pp. 5515-5532, Nov. 2010.

\bibitem{ulukus}
E. Ekrem and S. Ulukus, ``The secrecy capacity region of the Gaussian MIMO
multi-receiver wiretap channel'', \emph{IEEE Trans. Inform. Theory}, vol. 57, no. 4, pp. 2083-2114, Apr. 2011.

\bibitem{hassibi2}
F. Oggier and B. Hassibi, ``The secrecy capacity of the MIMO wiretap channel,'' \emph{IEEE Trans. Inform. Theory.}, vol. 57, no. 8, pp.4961-4972, Aug. 2011.

\bibitem{GEYRC12}
G. Geraci, M. Egan, J. Yuan, A. Razi, and I. Collings, ``Secrecy sum-rates for multi-user MIMO regularized channel inversion precoding,'' \emph{IEEE Trans. Commun.}, vol. 60, no. 11, pp. 3472-3482, Nov. 2012.

\bibitem{masksec}
M. Pei, J. Wei, K. -K. Wong, and X. Wang, ``Masked beamforming for multiuser MIMO wiretap channels with imperfect CSI,'' \emph{IEEE Trans. Wireless Commun.}, vol. 11, no. 2, pp. 544-549, Feb. 2012.

\bibitem{negi}
S. Goel and R. Negi, ``Guaranteeing secrecy using artificial noise,'' \emph{IEEE Trans. Wireless Commun.}, vol. 7, no. 6, pp. 2180-2189, June 2008.

\bibitem{zhou}
X. Zhou and M. R. McKay, ``Secure transmission with artificial noise
over fading channels: achievable rate and optimal power allocation,''
\emph{IEEE Trans. Veh. Technol.}, vol. 59, pp. 3831-3842, July 2010.

\bibitem{robsec}
A. Mukherjee, and A. L. Swindlehurst, ``Robust beamforming for security in MIMO wiretap channels with imperfect CSI,'' \emph{IEEE Trans. Sig. Proc.}, vol. 59, no. 1, pp. 351-361, Jan. 2011.

\bibitem{survey}
F. Rusek, D. Persson, B. K. Lau, E. G. Larsson, T. L. Marzetta, O. Edfors, and F. Tufvesson, ``Scaling up MIMO: Opportunities and challenges with very large arrays,'' \emph{IEEE Sig. Proc. Mag.}, vol. 30, no. 1, pp. 40-46, Jan. 2013.

\bibitem{noncooperative}
T. L. Marzetta, ``Noncooperative cellular wireless with unlimited numbers of BS antennas, '' \emph{IEEE Trans. Wireless Commun.}, vol. 9, no. 11, pp. 3590-3600, Nov. 2010.

\bibitem{energyspectraleff}
H. Q. Ngo, E. G. Larsson, and T. L. Marzetta, ``Energy and spectral efficiency of very large multiuser MIMO systems,'' \emph{IEEE Trans. Commun.}, vol. 61, no. 4, pp. 1436-1449, Apr. 2013.

\bibitem{training}
T. L. Marzetta, ``How many training is required for multiuser MIMO,'' in \emph{Proc. Fortieth Asilomar Conf. on Signals, Systems, and Computers (ACSSC'06)}, pp. 359-363, Pacific Grove, CA, Oct. 2006.

\bibitem{pilotcontam}
J. Jose, A. Ashikhmin, T. L. Marzetta, and S. Vishwanath, ``Pilot contamination and precoding in multi-cell TDD systems,'' \emph{IEEE Trans. Wireless Commun.}, vol. 10, no. 8, pp. 2640-2651, Aug. 2011.

%

\bibitem{ZMH12}
X. Zhou, B. Maham, and A. Hjorungnes, ``Pilot contamination for active evesdropping,'' \emph{IEEE Trans. Wireless Commun.}, vol. 11, no. 3, pp. 903-907, Mar. 2012.

\bibitem{KZWO13}
D. Kapetanovic, G. Zheng, K.-K. Wong, and B. Ottersten, ``Detection of pilot contamination attack using random training in massive MIMO,'' in \emph{Proc. IEEE Intern. Symp. Personal, Indoor and Mobile Radio Commun. (PIMRC)}, pp. 13-18, London, UK, Sept. 2013.

\bibitem{uldl}
J. Hoydis, S. ten Brink, and M. Debbah, ``Massive MIMO in UL/DL cellular systems: How many antennas do we need,'' \emph{IEEE J. Sel. Areas Commun}, vol. 31, no. 2, pp. 160-171, Feb. 2013.

\bibitem{Ian}
J. Zhang, R. W. Heath Jr., M. Koutouris, and J. G. Andrews, ``Mode
switching for MIMO broadcast channel based on delay and channel
quantization,'' \emph{EURASIP Journal on Advances in Signal Processing}, vol. 2009, doi:10.1155/2009/802548.

\bibitem{ud}
M. Kobayashi, N. Jindal, and G. Caire, ``Training and feedback optimization for multiuser MIMO downlink," \emph{IEEE Trans. Commun.}, vol. 59, no. 8, pp. 2228-2240, Aug. 2011.

\bibitem{CK78}
I. Csiszar and J. K\"orner, ``Broadcast channels with confidential messages," \emph{IEEE Trans. Inf. Theory}, vol. 24, no. 3, pp. 339-348, May 1978.

\bibitem{tse}
D. Tse and P. Viswanath, ``Fundemantals of wireless communications,''\emph{Cambridge University Press}, 2005.

\bibitem{gao}
H. Gao, P. J. Smith, and M. V. Clark, ``Theoretical reliability of MMSE
linear diversity combining in Rayleigh-fading additive interference
channels,'' \emph{IEEE Trans. Commun.}, vol. 46, no. 5, pp. 666-672, May 1998.

\bibitem{rmt}
A. M. Tulino and S. Verdu, ``Random matrix theory and wireless communications, '' \emph{Foundations and Trends in Communications and Information Theory}, vol. 1, no. 1, pp. 1-182, Jun. 2004.

\bibitem{appro}
Q. T. Zhang and D. P. Liu, ``A simple capacity formula for correlated diversity
Rician channels,'' \emph{IEEE Commun. Lett.}, vol. 6, no. 11, pp. 481-483, Nov. 2002.

\bibitem{wishart}
S. W. Nydick, ``The Wishart and Inverse Wishart Distributions,'' May 2012, [online] http://www.tc.umn.edu/~nydic001/docs/unpubs/Wishart Distribution.pdf.


\end{thebibliography}
\end{document}